\documentclass[useAMS,usenatbib]{mn2e}
\pdfoutput=1
\usepackage{natbib}
\usepackage{ graphicx}
\usepackage{ subfigure}
\usepackage{multirow}
\usepackage{appendix}

\title[Clumpy black hole accretion and feedback]{Black hole growth and AGN feedback under clumpy accretion}
\author[C. DeGraf et al.]  {C. DeGraf$^{1,2}$ A. Dekel$^{1}$, J. Gabor$^{3}$, F. Bournaud$^{3}$ \\
{1} {Center for Astrophysics and Planetary Science, Racah Institute of Physics, The Hebrew University,
 Jerusalem 91904 Israel} \\
{2} {Institute of Astronomy and Kavli Institute for Cosmology, University of Cambridge, Madingley Road, Cambridge, CB3 0HA, UK} \\
{3} {CEA-Saclay, 91190 Gif-sur-Yvette, France}}
\def\simgt{\lower.5ex\hbox{$\; \buildrel > \over \sim \;$}}

\begin{document}

\date{Submitted to MNRAS}
\pagerange{\pageref{firstpage}--\pageref{lastpage}}
\pubyear{20??}

\maketitle
\begin{abstract}
High-resolution simulations of supermassive black holes in isolated galaxies have suggested the importance of short ($\sim$10 Myr) episodes of rapid accretion caused by interactions between the black hole and massive dense clouds within the host.  Accretion of such clouds could potentially provide the dominant source for black hole growth in high-z galaxies, but it remains unresolved in cosmological simulations.  Using a stochastic subgrid model calibrated by high-resolution isolated galaxy simulations, we investigate the impact that variability in black hole accretion rates has on black hole growth and the evolution of the host galaxy.  We find this clumpy accretion to more efficiently fuel high-redshift black hole growth.  This increased mass allows for more rapid accretion even in the absence of high-density clumps, compounding the effect and resulting in substantially faster overall black hole growth.  This increased growth allows the black hole to efficiently evacuate gas from the central region of the galaxy, driving strong winds up to $\sim$2500 km/s, producing outflows $\sim$10x stronger than the smooth accretion case, suppressing the inflow of gas onto the host galaxy, and suppressing the star formation within the galaxy by as much as a factor of two.  This suggests that the proper incorporation of variability is a key factor in the co-evolution between black holes and their hosts.

\end{abstract}

\section{Introduction}
\label{sec:Introduction}

Observations suggest that supermassive black holes are to be found at the centers of most galaxies \citep{KormendyRichstone1995}, and properties of the black hole and the host galaxies are strongly correlated \citep{Magorrian1998, FerrareseMerritt2000, Gebhardt2000, Tremaine2002,Novak2006, GrahamDriver2007, Cattaneo2009, KormendyHo2013, McConnellMa2012}.  These correlations suggest that the growth of a black hole and the evolution of its host galaxy influence one another.  As such, black holes provide a means to better understand the evolution of galaxies, and may provide a key aspect to this evolution.  One of the most common explanations for this correlation is that quasar feedback from the central black hole may influence the host galaxy \citep[e.g.][]{BurkertSilk2001, Granato2004, Sazonov2004, Springel2005, Churazov2005, KawataGibson2005, DiMatteo2005, Bower2006, Begelman2006, Croton2006, Malbon2007, CiottiOstriker2007, Sijacki2007, Hopkins2007, Sijacki2009, DiMatteo2012, DeGraf2012, Dubois2013a, Dubois2013b}.  This feedback energy may be sufficient to unbind gas within the galaxy, driving strong outflows \citep{SilkRees1998, WyitheLoeb2003}.  Observations of galactic-scale outflows have been made \citep[e.g.][]{Fabian2006, Spoon2013, Veilleux2013, Cicone2014}, showing that such outflows certainly exist.  Furthermore, there is evidence that the strongest velocities are located in the central-most region of the galaxy \citep{Rupke2005, RupkeVeilleux2011}, possibly suggesting that the driving force behind them is indeed a centrally-located AGN rather than more widely-distributed feedback sources such as stars and supernovae.

Driving these large-scale outflows necessarily requires a large energy output from the AGN, which in turn requires a significant source of gas which can reach the black hole at the galactic center. The angular momentum loss required for this infall can pose a challenge.  One of the more commonly-posed explanations is that a gas-rich merger can drive gas toward the black hole.  Theoretical work suggests that mergers should drive significant AGN activity \citep[e.g.][]{Hernquist1989, DiMatteo2005, Hopkins2005d, Hopkins2005b, Hopkins2008, Johansson2009, Debuhr2010, Debuhr2011} and some observations support this \citep{Ellison2011}.  However, there have also been many studies which find that, although mergers may drive some AGN activity, the majority of AGN are found in isolated galaxies \citep{Schmitt2001, ColdwellLambas2006, Grogin2005, Georgakakis2009, Gabor2009, Cisternas2011, Kocevski2012}, suggesting that an alternate, secular mechanism may be the primary driving force in AGN activity.  One of the main drivers of AGN activity at high-redshift is believed to be cold flows: high-density streams of low angular momentum cold gas flowing along the cosmic web, whose high density and low temperature allow efficient penetration of halos to the innermost regions where they can continually fuel black hole growth \citep{DiMatteo2012, Dubois2012}.In addition, theoretical work has suggested that in high-z, gas-rich galaxies, violent disk instabilities can drive gas inflow and produce dense clumps of gas which can be driven in toward the galactic center \citep{Dekel2009b, Ceverino2010, Bournaud2011, Mandelker2014}, which may be a primary cause of AGN activity \citep{Bournaud2012} and provides a means of rapidly growing black holes even in the absence of cold streams.

In a companion paper, \citet{GaborBournaud2013} used high (6 pc) resolution simulations to show that accretion onto black holes in gas-rich galaxies can be highly variable, with strong bursts of accretion caused by dense infalling gas clouds.  These accretion events were found to generate strong outflows, but without significant effect on the host galaxy \citep{GaborBournaud2014}, at least over short ($\sim 100$ Myr) timescales and in the absence of cosmological gas flows and mergers.  Similarly, \citet{NovakOstrikerCiotti2011} used 2D simulations to show that cool shells of gas would fragment, with the fragmentation leading to bursts of accretion and an overall higher accretion rate.  In this paper we investigate the impact of periodic bursts of accretion on the growth of black holes and the corresponding effect they have on the host galaxy in a cosmological context, in which the black holes grow by several orders of magnitude (spanning both quiescent AGN phases and stronger quasar phases of extended Eddington growth).  We use zoom-in simulations to achieve $\sim 100$ pc resolution for galaxies in a cosmological environment, utilizing a stochastic subgrid model to incorporate the accretion of unresolved high-density gas clouds.  
We investigate how, in the context of cosmological gas inflow and galaxy mergers, the inclusion of periodic, high-accretion events affects black hole growth, and the impact this has on the host galaxy morphology and star formation rate, and on galactic gas inflow and outflow.

The paper is organized as follows: In Section \ref{sec:Method} we describe the simulations used and detail the subgrid model for the periodic accretion bursts.  In Section \ref{sec:bhgrowth} we investigate the impact of these periodic accretion bursts on black hole growth.  In Section \ref{sec:host} we show how AGN feedback from these accretion bursts can affect the host, specifically host morphology (\ref{sec:hostmorphology}), gas properties of the host (\ref{sec:gas_impact}), and gas inflows/outflows (\ref{sec:inflow_outflow}).  In Section \ref{sec:earlytime} we compare the impact at earlier times, providing a more direct comparison to the high-resolution isolated galaxy run.  Finally, we summarize our results in Section \ref{sec:Conclusions}.

\section{Method}
\label{sec:Method}

\subsection{RAMSES Code}
\label{sec:RAMSES}

For this work we ran cosmological zoom-in simulations using the Adaptive Mesh Refinement (AMR) code RAMSES \citep{Teyssier2002}, which uses particles (acting as collisionless fluid) to model dark matter and stars, while gas is modeled by solving the hydrodynamic equations on a cubic grid of cells which vary in size.  This code incorporates cooling, star formation, stellar feedback, and black holes.  Cooling is performed as a sink term in the thermal energy of the gas.  We allow gas to cool to a minimum temperature floor of $T_{\rm{th}}=10^4$ K, together with a density-dependent temperature floor to keep the local Jeans length above 4 cell-sizes, with a polytropic equation of state $T=T_{\rm{th}} (\rho/\rho_{\rm{th}})^{\gamma-1}$ with $\gamma = 2$, thereby preventing artificial fragmentation \citep[see, e.g.,][]{Truelove1997}. A uniform UV background (neglecting local sources) heats the gas according to the model of \citet{HaardtMadau1996}, using $z_{\rm{reion}} = 8.5$.

Star formation is performed in gas cells above the critical density $n_H > 0.1 \: \rm{cm}^{-3}$.  The star formation rate is $\dot{\rho} = \epsilon_* \rho_{\rm{gas}}/t_{ff}$, where $\rho_{\rm{gas}}$ is the gas density in the cell, $t_{ff} = (3 \pi/32G\rho_{\rm{gas}})^{1/2}$ is the local free-fall time of the gas, and $\epsilon_* = 0.01$ is the star formation efficiency \citep{Kennicutt1998, KrumholzTan2007}.  New star particles are then formed stochastically according to the star formation rate of the cell \citep{RaseraTeyssier2006}, initially given the position and velocity of the host cell, but uncoupled from the cell.  Supernova feedback is modeled by depositing $10\%$ of a star particles initial mass into the local cell 10 Myr after formation.  The energy released is $10^{50} \rm{erg}/M_\odot$ per $M_\odot$ of stars which go supernova. The energy is deposited thermally onto the gas, and cooling within the cell in which the energy is deposited is delayed by 2 Myr to prevent overcooling of the gas \citep[following approachs taken by, e.g.,][]{Stinson2006, Teyssier2013, GaborBournaud2014}.  

We use the same supermassive black hole prescription as \citet{GaborBournaud2013} \citep[see also][]{Dubois2012}.  Black holes are represented as sink particles, seeded into cells whose densities surpass $n_H > 1 \rm{cm}^{-3}$, with an initial mass of $M_{\rm{seed}}=10^5 M_\odot$.  Rather than representing the initial formation of an unresolved seed, this mass is broadly consistent with multiple mechanisms for seed formation, e.g. collapse of PopIII stars \citep[e.g.][]{BrommLarson2004, Yoshida2006} or direct collapse of massive gas clouds \citep[e.g.][]{BrommLoeb2003, Begelman2006}, followed by sufficient growth to reach $M_{\rm{seed}}$.  We also prevent black holes from forming within 25 kpc of another BH, thereby preventing multiple BHs from forming within the same galaxy.  Once seeded, the black hole grows through gas accretion and BH-BH mergers.  Gas accretion is modeled as 
\begin{equation}
\dot{M}_{\rm{BH}}=(4 \alpha \pi G^2 M_{\rm{BH}}^2 \rho)/(c_s^2+v_{\rm{rel}}^2)^{3/2}
\label{eqn:bondi}
\end{equation}
\citep{HoyleLyttleton1939, BondiHoyle1944, Bondi1952}, where $\rho$ is the gas density, $c_s$ is the sound speed of the gas, $v_{\rm{rel}}$ is the velocity of the black hole relative to the gas (calculated within a sphere of 4$r_{\rm{min}}$, where $r_{\rm{min}}$ is the minimum resolution element of the simulation), and $\alpha = (\rho/\rho_0)^2$ for $\rho > \rho_0$ and $\alpha=1$ for $\rho < \rho_0$ \citep{BoothSchaye2009}.  To prevent unphysically high accretion rates, we cap $\dot{M}_{\rm{BH}}$ at the Eddington limit 
\begin{equation}
\dot{M}_{\rm{edd}} = (4 \pi G M_{\rm{BH}} m_p)/(\epsilon_r \sigma_T c)
\label{eqn:edd}
\end{equation}
 \citep{Eddington1916}, where $m_p$ is the mass of a proton, $\sigma_T$ is the Thomson scattering cross section, $c$ is the speed of light, and $\epsilon_r$ is the radiative efficiency for the accreting gas, assumed to be 0.1 \citep{ShakuraSunyaev1973}.

Black hole feedback is accomplished using a thermal feedback model, depositing $\dot{E}_{\rm{BH}} = \epsilon_f \epsilon_r \dot{M}_{\rm{BH}}$ \citep[$\epsilon_f = 0.15$ is the feedback efficiency, selected to reproduce the scaling relations between the black hole and the host galaxy, see][]{Dubois2012} onto the gas within $4 r_{\rm{min}}$ of the BH. To prevent instantaneous overcooling of the gas (which will tend to happen at low temperatures where the cooling rate is high enough), we only deposit this energy if it is sufficient to heat the gas to at least $10^7$ K, otherwise the energy is stored until this threshold can be reached \citep{BoothSchaye2009}.  To prevent unphysically high temperatures, if the thermal feedback is sufficient to heat the gas in excess of $5 \times 10^9$ K, the injection region is iteratively expanded until the resulting temperature will be below this level.  

\subsection{Clumpy Accretion model}
\label{sec:clumpymodel}

The key modification to the black hole treatment in this paper is the incorporation of unresolved high-density gas clouds.  

\begin{figure}
\centering
\includegraphics[width=8cm]{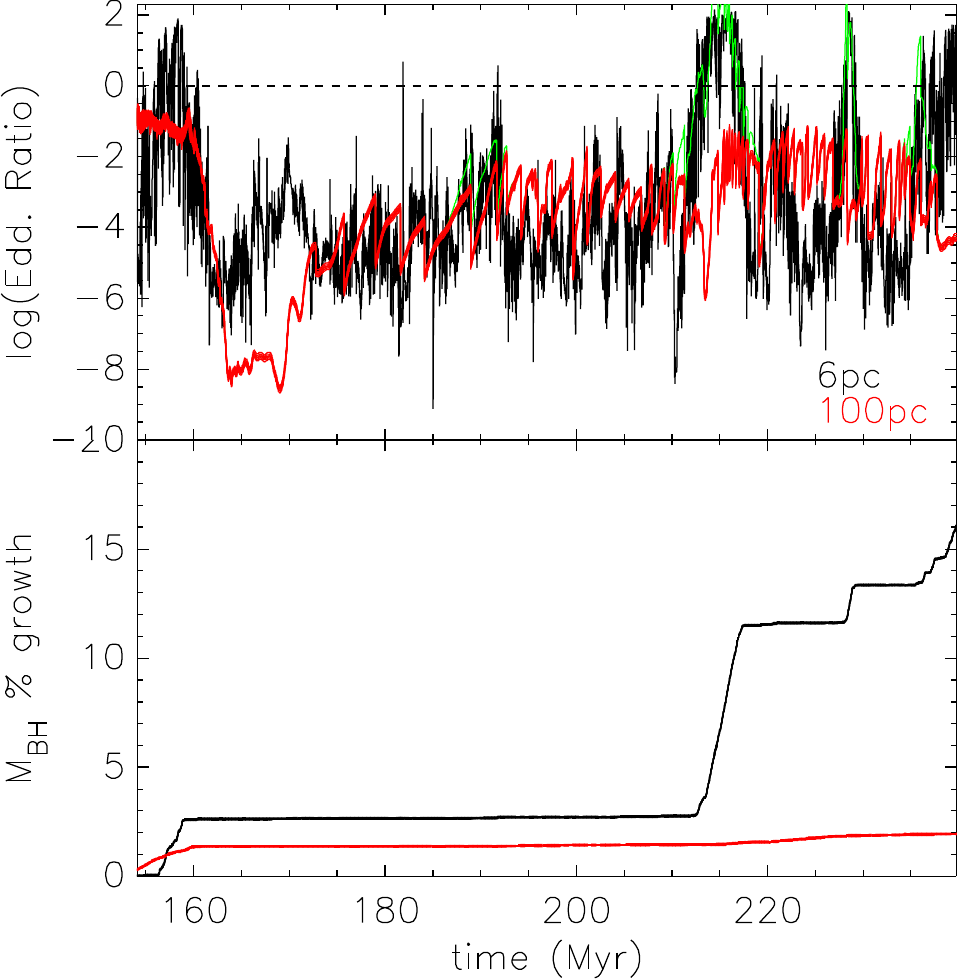}
\caption{Resolution dependence of black hole growth.  \textit{Top:} Eddington fraction ($\dot{M}_{\rm{BH}}/\dot{M}_{\rm{edd}}$) for black hole in isolated galaxy simulation using 6pc resolution (black) and 100pc resolution (red).  Dashed line shows the Eddington limit. Green curve shows the 100pc run with added gaussian curves for bursts of accretion (used for stochastic subgrid model, see Section \ref{sec:clumpymodel}. \textit{Bottom:} Black hole growth (as a percentage of its initial mass) over the course of the simulations.  Lowering the resolution smooths out the highest peaks and leads to significantly lower BH growth.}
\label{fig:resolutioncompare}
\end{figure}

\citet{GaborBournaud2013} found that the accretion of high-density clouds of gas could be the dominant factor is black hole growth (at least among gas-rich, high-z galaxies), based on isolated galaxy simulations with 6pc resolution.  Because these clouds of gas are only $\sim 100-300$ pc in radius, they remain unresolved in the majority of cosmological simulations.  To investigate the effect of using resolution more typical for cosmological runs, we re-ran the M4f50 run from \citet{GaborBournaud2013} at the lower resolution of 100pc.  In Figure \ref{fig:resolutioncompare} we show the comparison between the 6pc resolution (black) and 100 pc resolution (red) runs.  On few-timestep scales, the high-resolution simulation exhibits more variation, as may be expected.  In addition to the general variability, we note two main differences.  First, the high-resolution simulation has several periods of high accretion on the order of 5-10 Myr.  Second, in the absence of these accretion events, the low-resolution simulation tends to accrete more rapidly, by a factor of $\sim 2.5-3$.  This increased accretion in the low-resolution run is seen most clearly in the upper panel of Figure \ref{fig:resolutioncompare} between 190-210 Myr and 220-235 Myr, where the red (low resolution) is clearly significantly higher than the black (high resolution), except during a clump accretion event.  In the bottom panel of Figure \ref{fig:resolutioncompare} we show the black hole growth for both the high- (black) and low- (red) resolution runs.  Here we see that the short accretion events contributing the majority of the black hole growth in the high-resolution simulation are missed in the low-resolution run, leaving the black hole at a much smaller final mass.  

Given the importance of these high-density gas clouds on the black hole growth, we incorporate a subgrid prescription to the accretion rate to boost the accretion as if a high-density gas cloud were able to be resolved.  We use a simple stochastic prescription for our model.  For any timestep in which a black hole is not already undergoing a burst of accretion, we allow for a new event to begin with a probability of $p_{\rm{burst}}$.  Each such event causes the accretion rate of the black hole to increase following a gaussian pattern, with a characteristic timescale ($\sigma_{\rm{burst}}$) and amplitude ($A_{\rm{burst}}$).  We use the high- and low- resolution runs (shown in Figure \ref{fig:resolutioncompare}) to calibrate the values of these parameters.  We do this by fitting Gaussians to the rate of increase in the ratio between the accretion rates of the two simulations, finding four events which occur during the comparison period.  From this, we incorporate four possible clump accretion events to our simulation, which each occur once in the 85 Myr high-resolution run.  These four events occur with amplitude $A_{\rm{burst}}=26.5, 6.22, 5.56, 4.66$, with timescales of $\sigma_{\rm{burst}}=1.83, 1.3, 0.4, 1.5$ Myr.  To account for the slower accretion during the smooth period (i.e. in the absence of a dense clump), the smooth accretion decreased by a factor of $\sim 2.6$ (matching the discrepancy in Figure \ref{fig:resolutioncompare}).  The model calibration is intended to give the lower resolution cosmological run a periodicity comparable to that of the high-resolution run that fully resolves high-density clouds.  Given the limited sample size of a single isolated galaxy over a relatively short timescale (for cosmological runs), this will not be a completely accurate parameterization, particularly since it does not depend on the various properties of the host.  In particular, we expect the formation of dense gas clumps to occur in gas rich galaxies, but such formation may be minimal (if at all) in gas-poor galaxies.  Indeed, this is found in \citet{GaborBournaud2013}, which explicitly showed that gas-poor galaxies did not form the dense clumps found in gas-rich galaxies.   For this paper, we investigate the impact that periodic accretion can have on the black hole growth and on the host in a galaxy where such clump formation occurs, for which this parameterization is sufficient.  We also note that the galaxy in our simulation has a gas fraction in excess of 50\% throughout the run, making the gas-rich run from \citet{GaborBournaud2013} the appropriate one to use here.  Having demonstrated the importance of including such variability in this work, we note that a full parameter study of how accretion of high-density clumps depends on host properties will be needed.  This will require a full suite of high-resolution isolated galaxy simulations to study clump formation as a function of galaxy properties, such as gas-fraction, galactic scale height, SFR, merger history, etc.  Such a suite of simulations is beyond the scope of this work, however, and is thus left for a followup project.  We also note the possibility that runs with resolution beyond the 6pc used in the isolated galaxy may exhibit slightly different behavior.  However, since the clumps in the isolated galaxy are well resolved (typically on the order of $\sim 100$ pc), we do not expect significant impact from higher resolution.

\subsection{Zoom-in simulations}
For this paper we run a set of zoom-in simulations within a $10$ Mpc box.  We assume a $\Lambda$CDM cosmology with cosmological parameters consistent with \citet{PlanckCosmological2013}: $\Omega_\Lambda = 0.68$, $\Omega_m=0.32$, $\Omega_b=0.05$, and $H_0$=67 km/s/Mpc.  Although these results are not fully consistent with the WMAP results \citep{Komatsu2011}, our investigation is based on a comparison of individual objects between two simulations runs, so our results should remain consistent regardless of the exact cosmology used.  Within the base 10 Mpc box, we resolve a zoom region of $\sim$1 Mpc about the largest black hole (based on a low-resolution test run), which reaches $\sim 10^7 M_\odot$ by $z=6$.  We resolve the zoom region to a maximum physical resolution of $\sim 76$pc, which corresponds to a refinement level of 17 at z=0.  The maximum refinement level at higher redshifts evolves with scale factor (maximum refinment level increase by 1 for each doubling of the scale factor), providing a maximum resolution that remains approximately constant with cosmic time.  Refinement is done when a cell contains more than 8 dark matter particles, or an equivalent gas mass.  Our minimum stellar particle mass is $\sim 4.3 * 10^{3} M_\odot$.  Given this base simulation setup, we run the same set of initial conditions \citep[generated with the \small{GRAFIC-2} subroutine, see][]{Bertschinger2001} using three versions of the code.  The ClumpyAccretion run includes our full black hole treatment, including the subgrid model for accretion of high-density clumps described in Section \ref{sec:clumpymodel}.  The SmoothAccretion run includes black holes, but using the standard accretion model described in Section \ref{sec:RAMSES}.  Note that we refer to this model as the SmoothAccretion since it lacks the periodic bursts of accretion caused by unresolved gas clouds, but the black hole accretion rate nonetheless varies based on the resolved gas properties around it.  Finally, the NoBH run is the base run which does not include black holes at all.  The primary analysis of all runs was performed with the data analysis toolkit \small{yt} \citep{YT2011}.

\section{Black Hole Growth}
\label{sec:bhgrowth}

\begin{figure}
\centering
\includegraphics[width=8cm]{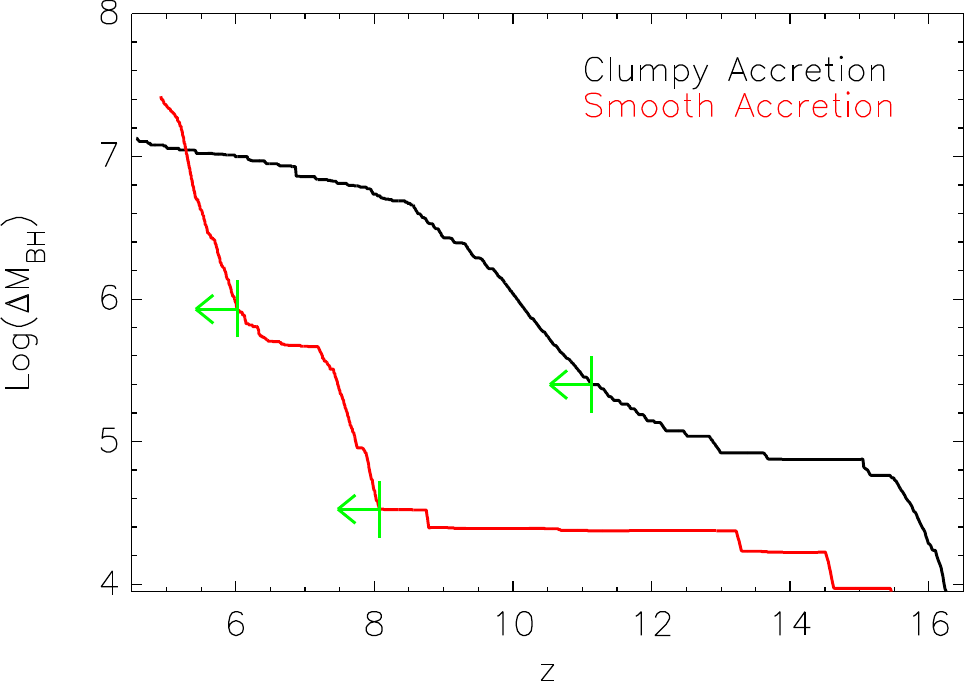}
\caption{Accreted mass onto our primary black hole in the clumpy-accretion model (black) and smooth-accretion model (red). Green arrows mark the onset of an extended Eddington regime.  Black hole mass builds up earlier in the clumpy accretion case, but also leads to a lower final mass.}
\label{fig:bh_growth}
\end{figure}

\begin{figure}
\centering
\includegraphics[width=8cm]{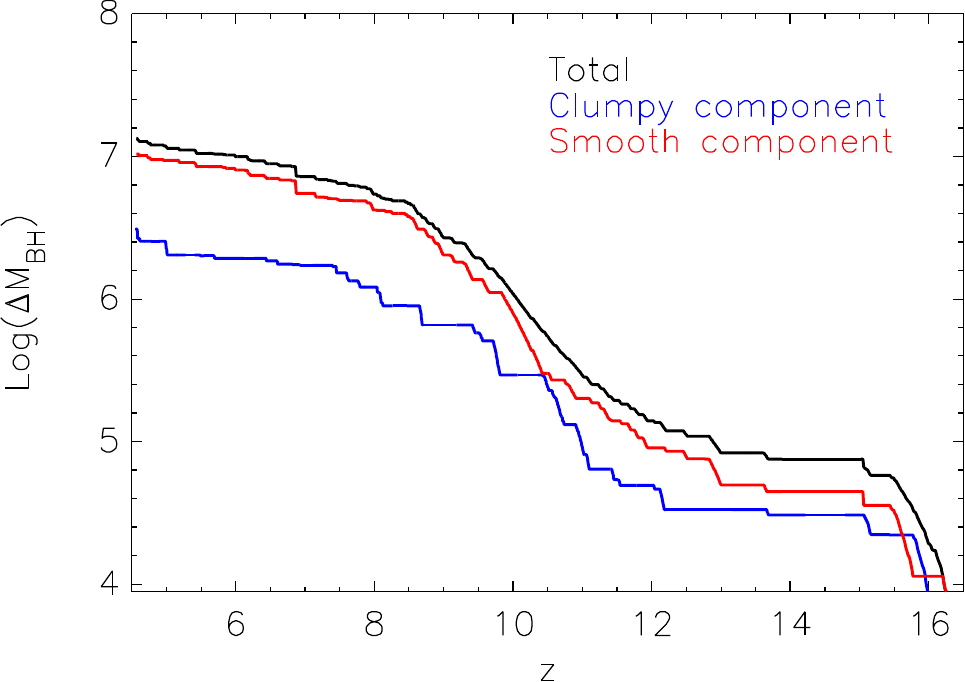}
\caption{The growth of our primary black hole in the clumpy-accretion model, showing the contribution to the accreted mass from accretion of dense clumps (red) and smooth infall between clump events (blue).  The relative importance of the clumpy component of accretion is strongest just as the black hole reaches Eddington at z $\sim$11.}
\label{fig:clumpy_growth_representation}
\end{figure}

In Figure \ref{fig:bh_growth} we show the accreted mass growth of our primary black hole in both the clumpy-accretion (black) and the smooth-accretion (red) runs, clearly showing a dramatically different growth history.  In both simulations, the black hole follows the typical growth behavior found in cosmological simluations \citep[e.g.][]{DiMatteo2008, DeGrafBHgrowth2012}: it undergoes an initial sub-Eddington growth phase, followed by an extended period of Eddington growth, and upon reaching a high enough mass (relative to its host), self-regulation kicks in, dramatically slowing the growth of the black hole.  The main difference between the runs is the onset time of the Eddington growth phase, which occurs much sooner in the clumpy-accretion model.  In the smooth-accretion model, the sub-Eddington phase is very long-lasting.  Without the added accretion from the dense clumps of gas, the black hole takes until $z \sim 8$ to grow massive enough to reach the Eddington regime.  In contrast, the clumpy-accretion model reaches the Eddington regime around $z \sim 10-11$, and has already reached the self-regulation regime by $z \sim 8$.  This substantial difference is due to the periods of clump accretion providing short time-scale bursts of Eddington accretion during the sub-Eddington regime.  In Figure \ref{fig:clumpy_growth_representation} we divide the total accreted matter (black) from the clumpy accretion run into two components: the accreted mass during clump-accretion events (blue) and in the absence of clumps (i.e. during smooth accretion; red).  

From these curves it appears that the accretion of clumps plays a relatively minor role.  However, this conclusion neglects two important factors.  First, the total mass gained during clump accretion is not a meaningful quantity, since the majority of growth occurs during the extended Eddington phase.  During this phase, the growth is capped at $M_{\rm{Edd}}$, and thus an incoming clump will not provide any increase in the accretion rate.  For this reason, the meaningful quantity to consider is the mass gained via clump accretion prior to the onset of Eddington accretion.  Based on this, we see that the black hole has gained approximately half of its mass through clump accretion near the onset of the Eddington phase (z $\sim$11), demonstrating a significant impact.  Even this check underestimates the importance of the clumps, however, as it neglects the exponential nature of the black hole growth.  Because the smooth accretion phases depend upon $M_{\rm{BH}}^2$, modest increases in mass at early times (such as those caused by early clump accretion events) have an exponential impact on the continued growth of the black hole, which is what causes the dramatic differences between the two simulations in Figure \ref{fig:bh_growth}.  Thus we note that relatively minor differences at very early times can significantly affect the late-time behavior of the black hole.

This ability to drive rapid growth at early times may be of significant importance to the seeding mechanisms for supermassive black holes.  Using a standard Bondi-like accretion rate, a very low-mass black hole (e.g. a $10^2 M_\odot$ seed from a PopIII star) will tend to accrete relatively slowly.  This can present a problem when attempting to reach the high masses seen in observations \citep[such at the $10^9 M_\odot$ BH found at $z \sim 7$ by ][]{Mortlock2011}.  However, the bursts of accretion provided by high-density clouds can produce substantially more rapid growth among small, early BH seeds.  Initial tests suggest that black holes seeded at masses of $\sim 10^3 M_\odot$ can still grow to $\sim 10^7 M_\odot$ by $z \sim 7$, which will provide more flexibility in the seeding prescriptions used in cosmological simulations.  

Furthermore, the early growth of a black hole can be highly sensitive to the seeding prescription, particularly the seed mass.  Although the final mass (maintained via self-regulation) may be relatively insensitive to the seeding prescription, the evolution to that final mass may be significantly different.  As Figure \ref{fig:bh_growth} shows, a larger mass early on can result in much faster overall growth.  For example, using a seed of $5 \times 10^4 M_\odot$ will take $\sim 2.5$ times longer to reach $10^{5.5} M_\odot$ ($\sim$ when our BH reaches the Eddington regime) than a seed of $10^5 M_\odot$ if we assume Bondi accretion with constant gas properties.  However, the clumpy-accretion events tend to occur at the Eddington rate, which depends on $M_{\rm{BH}}$ rather than $M_{\rm{BH}}^2$ (see Equations \ref{eqn:bondi} and \ref{eqn:edd}), making it much less sensitive to the seed mass.  If we assume all the growth is from these bursts at Eddington, the $5 \times 10^4 M_\odot$ seed will only take 1.6 times longer than the $10^5 M_\odot$ seed.  Although the actual result will be somewhere between these expectations (and also depend on the evolution of the gas properties), this clearly shows that the incorporation of clumpy accretion has the potential to make the black hole growth much less sensitive to the seed mass.  A full study of the impact of clumpy accretion on black hole seeding prescriptions is beyond the scope of this paper, but may prove useful for studies attempting to isolate the formation mechanism for supermassive black hole seeds.

\section{Impact on host}
\label{sec:host}
\subsection{Morphology}
\label{sec:hostmorphology}

\begin{figure*}
\centering
\includegraphics[width=17cm]{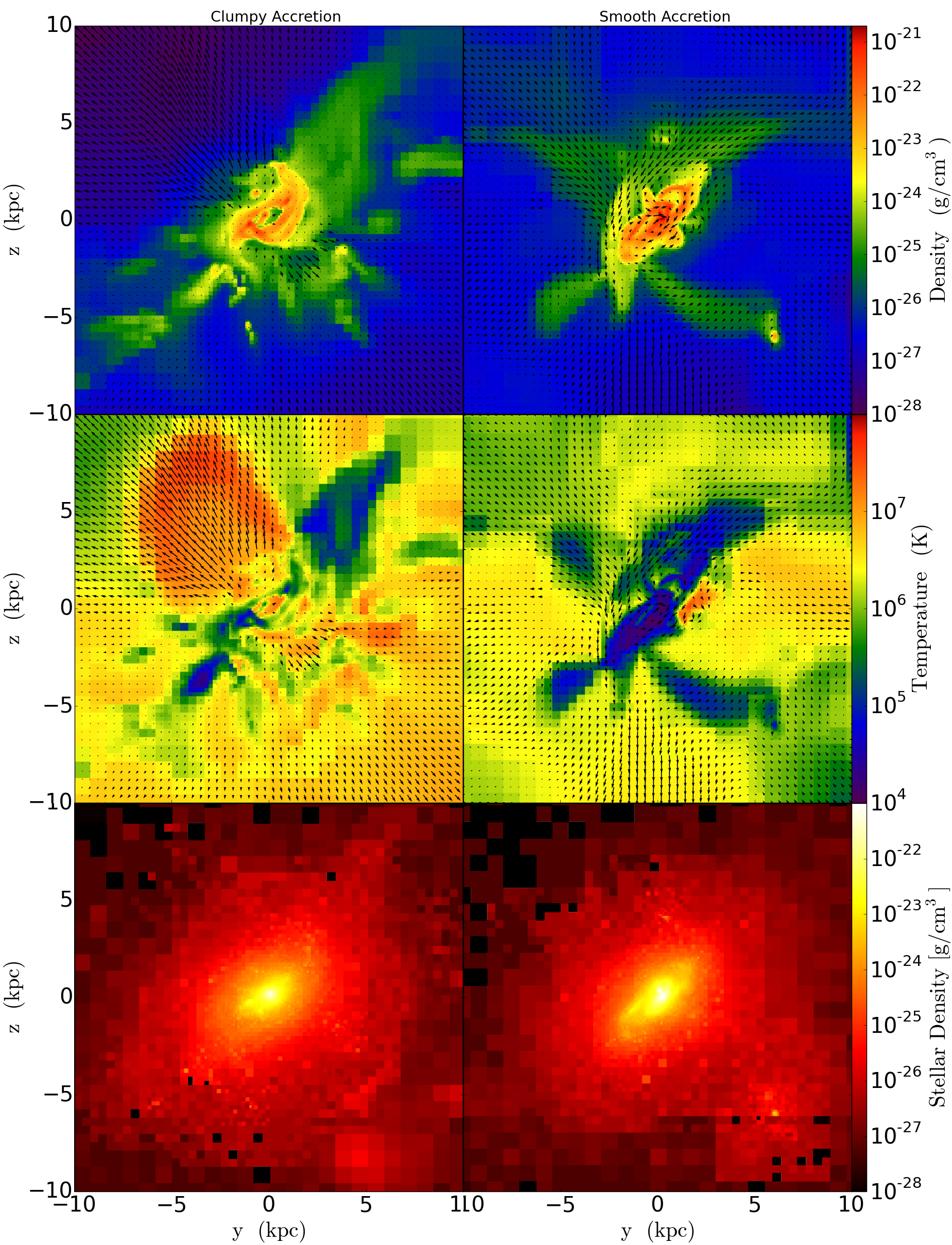}
\caption{Projected maps of our simulated galaxy at z=7.65 showing AGN-driven outflows in the clumpy-accretion model.  \textit{Top:} Gas density; \textit{Middle:} Gas temperature; \textit{Bottom:} Stellar Density.  Left panels show the galaxy in the clumpy-accretion simulation; Right panels show the galaxy in the smooth-accretion simulation.  Each plot is produced from a slice 6-kpc thick.  The clumpy accretion has evacuated the center-most region of gas, and drives rapid outflows of hot gas.}
\label{fig:projection_67}
\end{figure*}

\begin{figure}
\centering
\includegraphics[width=9cm]{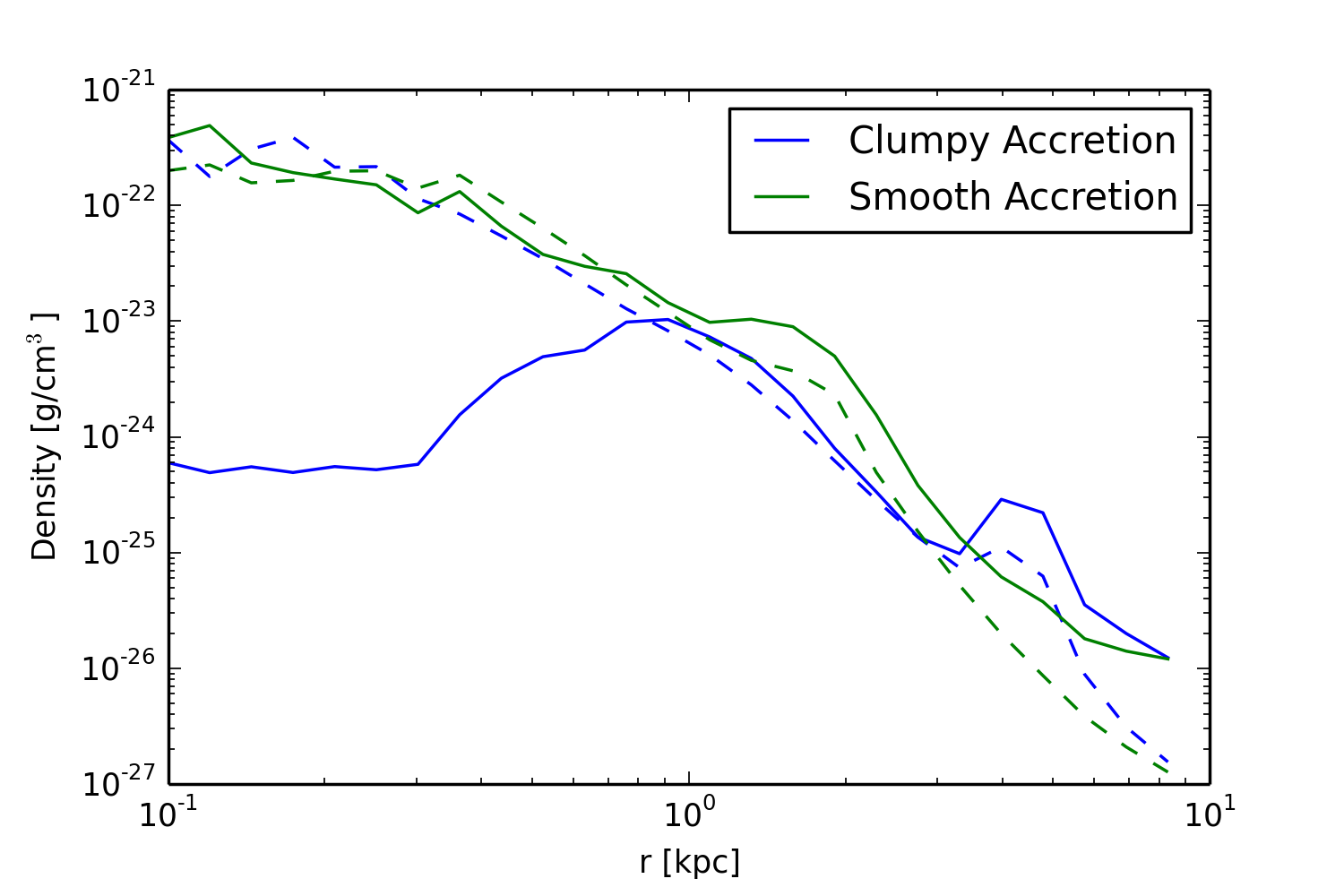}
\caption{Density profiles for the clumpy accretion model (blue) and smooth accretion model (green) at z=7.65.  \textit{Solid lines} - gas; \textit{Dashed lines} - stars.  Clumpy accretion triggers AGN feedback that lowers the nuclear gas density compared to the smooth accretion case.  The stellar profile is minimally affected, with the smooth accretion model having slightly more stars than the clumpy accretion model.}
\label{fig:density_profile_comparison}
\end{figure}

In Figure \ref{fig:projection_67} we show images of the gas density (top), gas temperature (middle), and stellar density (bottom) of our galaxy in both the clumpy-accretion model (left) and the smooth-accretion model (right).  This figure shows the qualitative effect that the clumpy-accretion model has on the environment in terms of general morphology, AGN-driven outflows, and effect on inflowing gas.  The redshift was selected to highlight an outlflow process, but we note that other redshifts after the extended Eddington phase are qualitatively similar (see Section \ref{sec:earlytime} for early time comparison).  In the density projections, the smooth-accretion model shows a well-defined disk of cold gas has formed without being disrupted.  The clumpy-accretion model, however, shows a less well-defined disk that is relatively puffed out in all directions, i.e. has a less-well defined disk plane.  More striking than this, however, is the central region of the galaxy, which has been evacuated of dense gas, leaving a substantial void of low-density, high temperature gas surrounding the black hole.  This is more clearly seen in Figure \ref{fig:density_profile_comparison}, which shows the gas density profile (solid lines) for the galaxy in both simulations.  These profiles show comparable densities above $\sim 1$ kpc (though slightly lower density in the clump-accretion model), but a dramatic difference (up to 2 dex) at smaller scales.  Note that the highest resolution cells are $\sim 0.1$ kpc, so the results at the smallest scales are not well-resolved, but the decrease at sub-kpc scales is well within the resolution of the simulation.  This clearly demonstrates the ability of the clump-fed AGN to evacuate the gas from the central region of the galaxy, which will necessarily lead to the suppression of the black hole growth (i.e. self-regulation) as well as quench star formation (investigated in more detail in Section \ref{sec:inflow_outflow}).

The temperature maps in Figure \ref{fig:projection_67} also show significant differences, with the clumpy-accretion model showing a hot region surrounding the black hole ($\sim 1$ kpc, corresponding to the evacuated region), outside of which there are regions of hot and cold gas.  In contrast, the disk in the smooth-accretion model remains cool with fewer regions of temperature variation.  Outside the galaxy, the clumpy-accretion model produces bubbles of hot gas inflating away from the black hole (similar to radio cavities observed in galaxies), showing clear evidence of AGN-driven outflows.  These hot bubbles of outflowing gas are completely lacking in the smooth-accretion model.  Consistent with the higher-resolution runs of \citet{GaborBournaud2014}, despite using a purely isotropic feedback model the outflows are nearly entirely out-of-plane, though they are not necessarily axisymmetric (see Section \ref{sec:in_vs_out_of_plane} for more details).  This anisotropy is purely a result of the local environment, with dense in-plane gas shielding the rest of the disk from the feedback energy, while the relatively low-density out-of-plane gas is effectively driven out.  Figure \ref{fig:projection_67} clearly shows the outflows driven almost exclusively in directions of low-density gas, and also shows that resolved cold, dense clumps effectively block the outflows.

Unlike the gas density and temperature, the stellar morphology is only weakly affected by the clumpy accretion model.  In the bottom panels of Figure \ref{fig:projection_67} we plot the stellar density maps, which show only minimal difference between the two runs.  The stars in the smooth accretion case are slightly flatter/more elongated than in the clumpy case, which has a more rounded stellar component.  This is consistent with the general gas distribution (top panels), and is a fairly small effect.  More significantly, we see no evidence of the evacuated region at the center of the galaxy.  This is confirmed in Figure \ref{fig:density_profile_comparison}, where the dashed lines show the stellar density profile.  We find the smooth accretion model has slightly higher stellar densities, but otherwise the \textit{distribution} of stars is largely unaffected by the AGN feedback, down to the smallest scales.  Thus we find, as expected, that the AGN feedback can have a strong impact on the gas, but has no direct affect on the stellar distribution.  It can \textit{indirectly} affect the galaxy's stellar mass by suppressing star formation (resuling in the slightly higher stellar densities in Figure \ref{fig:density_profile_comparison}), which we investigate in more detail in Section \ref{sec:inflow_outflow}.

\subsection{Gas properties}
\label{sec:gas_impact}

\begin{figure*}
\centering
\includegraphics[width=18cm]{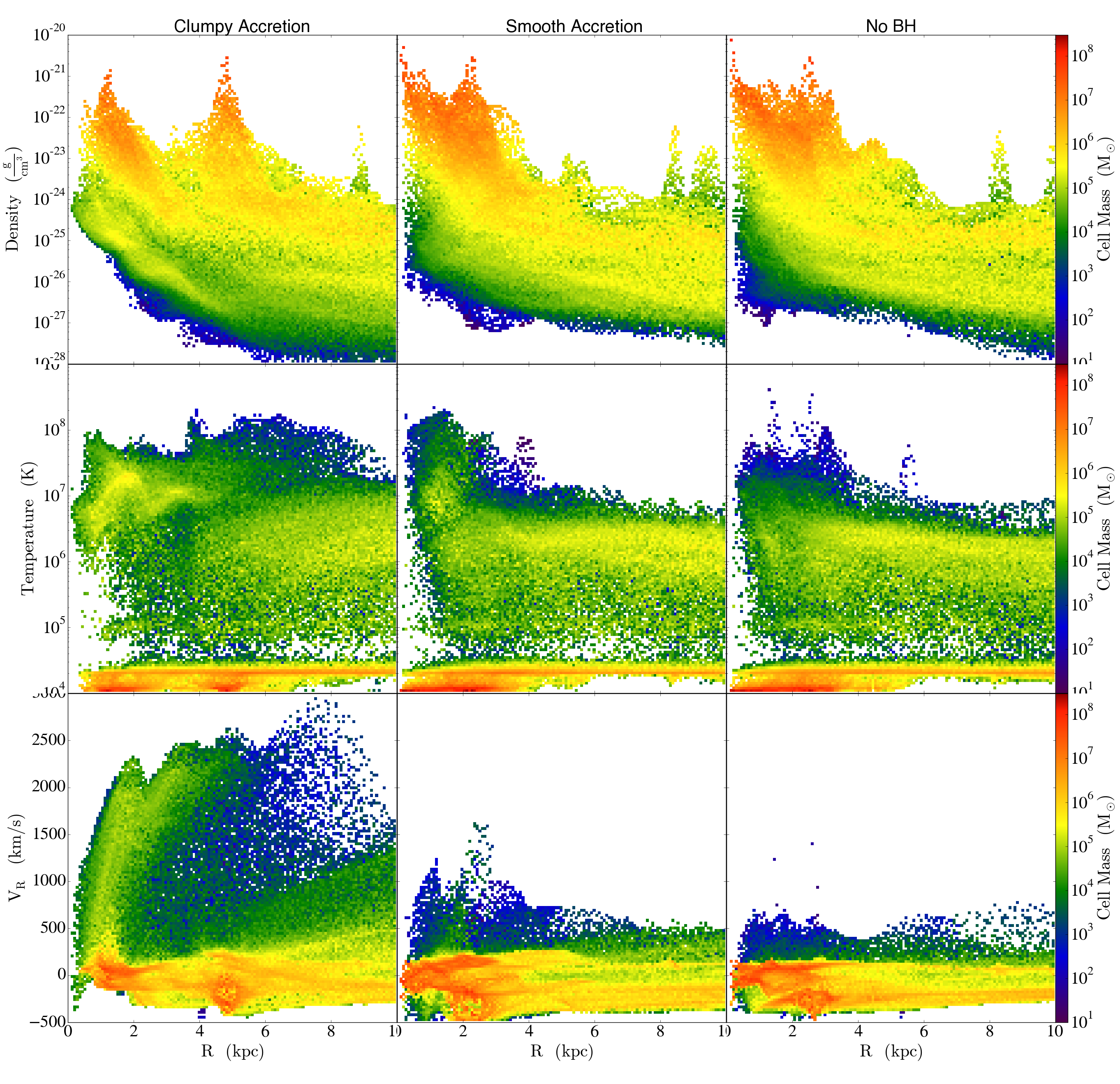}
\caption{Gas properties of the host galaxy at z=7.65: the density (top), temperature (middle), and radial velocity (bottom) of the gas as a function of radial distance from the galaxy center for the clumpy-accretion (left), smooth-accretion (middle), and no-bh (right) simulations.  Clumpy accretion triggers feedback that heats nuclear gas and drives high-velocity outflows not seen in the smooth accretion or no-bh simulations.}
\label{fig:radial_properties_67}
\end{figure*}

In addition to the general morphology, we find noteable differences in the gas properties within the host galaxy.  In Figure \ref{fig:radial_properties_67} we show the distribution of gas density (top), temperature (middle), and radial velocity (bottom) vs. distance from the galaxy center for all three simulation runs at z$\sim$7.65, matching Figure \ref{fig:projection_67}.  Pixel color represents the mass of the gas at the given pixel.  First, we note that the difference between the smooth-accretion and no-bh runs is quite small.  The smooth-accretion black hole heats some of the nearby ($< \: \sim$3 kpc) gas to higher temperatures than the no-bh case, and there is some outflowing gas driven at slightly higher velocities, but they are otherwise qualitatively similar.  The clumpy-accretion model, however, is substantially different.  In the density distribution, we see that in the vicinity of the black hole, the very low-density gas ($\sim 10^{-25}-10^{-26} \rm{g/cm}^3$ within $\sim$2 kpc, at the bottom left of the panel) has been completely removed in the clumpy-accretion run.  At larger radii, this run has extremely low-density gas ($< 10^{-27}$) which is completely missing in the smooth-accretion run.  This suggests that the bulk of the low-density gas near the black hole was driven away as outflows, and is thus found at larger radii.  

The temperature distributions in Figure \ref{fig:radial_properties_67} show a similar picture.  Although the bulk of the very cold (and high-density) gas remains, the majority of inner ($< 2$ kpc) cool gas (between $3 \times 10^4$ and $10^6$ K) has been heated to higher temperatures, and there is significantly more hot gas ($> 10^7$ K) at larger radii.  This is consistent with the general picture that the nearby gas has been heated to high temperature and driven out to larger radii.  In the bottom panel we confirm this high-velocity gas outflow driven by the clumpy-accretion black hole, with high velocities (up to 3000 km/s) maintained out to radii of 8 kpc, compared to the smooth-accretion model where almost no gas exceeds 500 km/s.  

\begin{figure*}
\centering
\includegraphics[width=18cm]{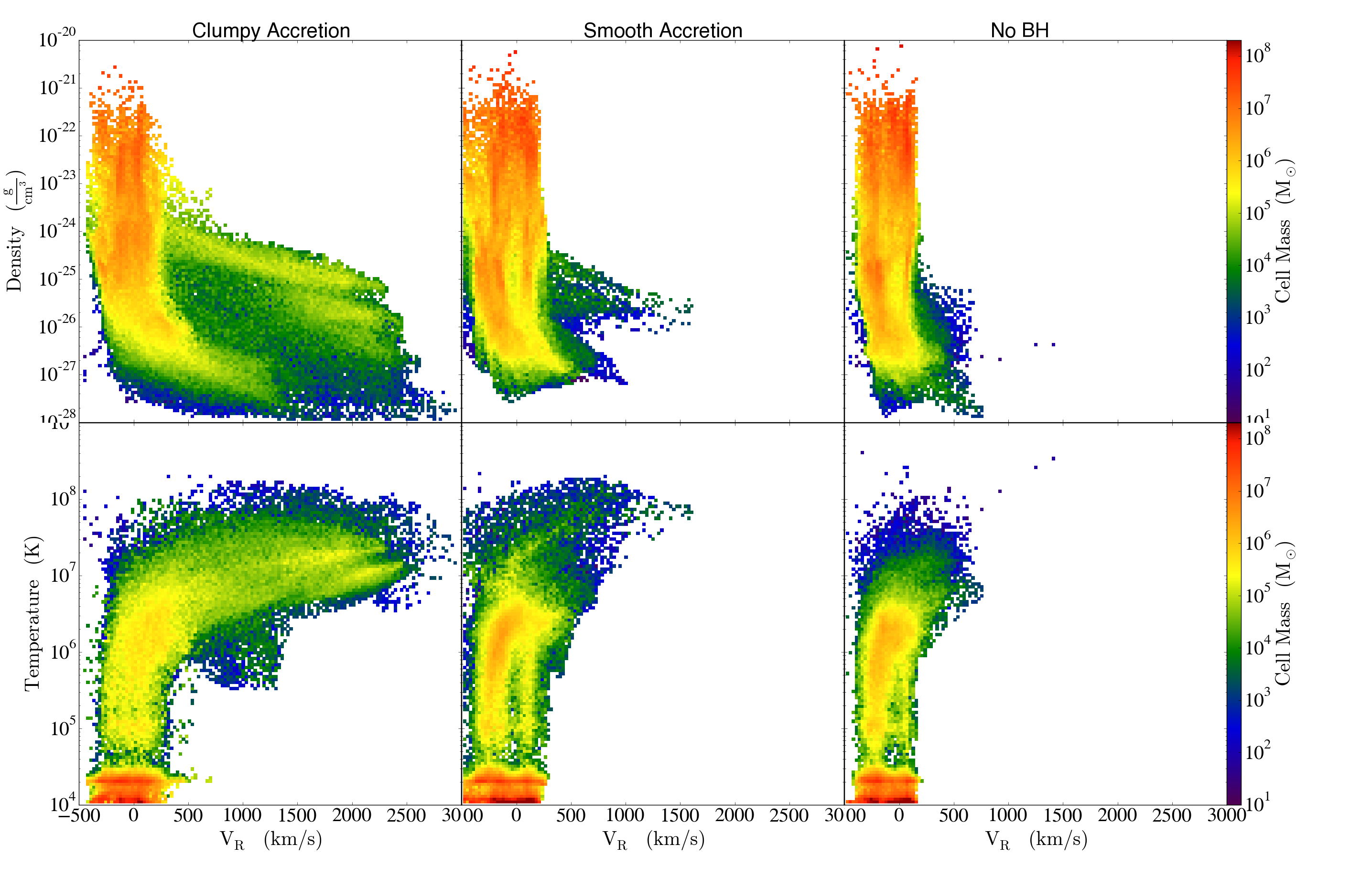}
\caption{Gas properties of the host galaxy at z=7.65: the density (top) and temperature (bottom) of the gas as a function of radial velocity for the clumpy-accretion (left), smooth-accretion (middle), and no-bh (right) simulations.  Clumpy accretion triggers hot, diffuse, high-velocity winds largely absent from the other simulations.}
\label{fig:velocity_properties_67}
\end{figure*}

Figure \ref{fig:velocity_properties_67} shows the distribution of gas densities and temperatures as a function of radial velocity.  Here we can explicitly see that the strongly outflowing gas found in the clumpy-accretion simulation is low-density ($< 10^{-24} g/cm^{3}$) and high temperature ($> 3 \times 10^6$ K, and most above $10^7$ K).  This is consistent with the high-resolution isolated galaxies of \citet{GaborBournaud2014}, who similarly found outflows consisting of hot, diffuse gas.  Since none of the strongly outflowing gas is at high densities, we deduce that the AGN driven outflows do not directly evacuate the starforming gas, which is dense.  Nevertheless, there are other means by which the AGN can suppress star formation, which we investigate further in the next section.

\subsection{Inflow and outflow rates}
\label{sec:inflow_outflow}

\begin{figure}
\centering
\includegraphics[width=8cm]{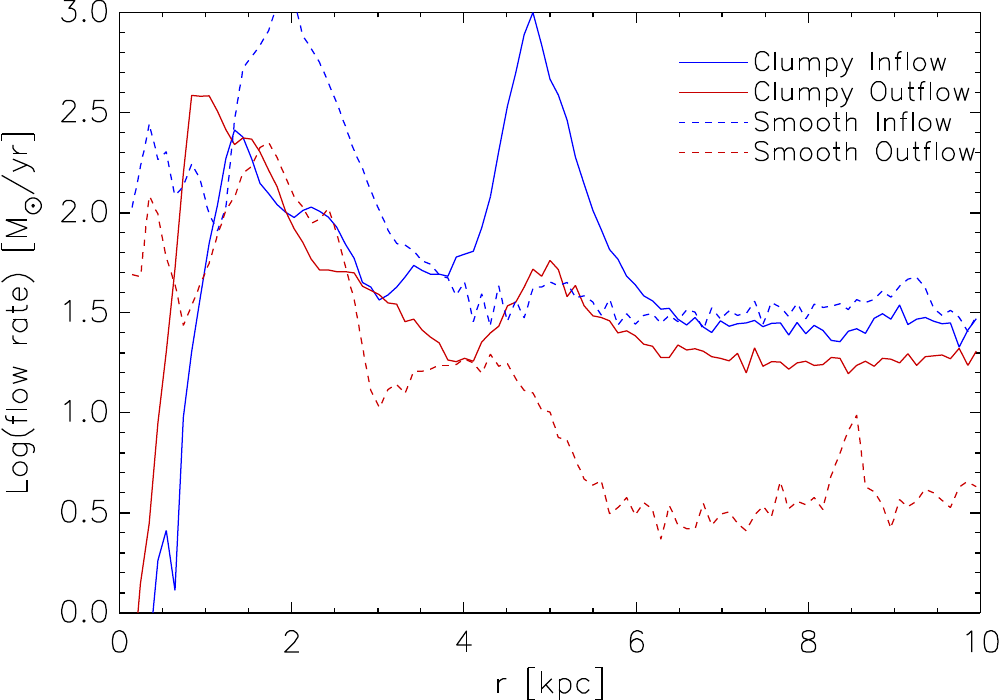}
\caption{Gas inflow (blue) and outflow (red) rates at z=7.65 as a function of radial distance from the black hole.  Clumpy accretion prevents flow into the innermost kpc and drives much stronger outflows out to large scales.  }
\label{fig:radial_flow_rates_67}
\end{figure}

In Figure \ref{fig:radial_flow_rates_67} we show the instantaneous gas inflow and outflow rates through spherical shells about the galaxy center.  These flow rates are calculated by $\dot{M}= \frac{1}{\Delta x} \sum{m_i v_i}$, where $m_i$, $v_i$ are the mass and radial velocity for each cell $i$ in the spherical shell, and $\Delta x$ is the shell thickness.  For thin shells, this is a reasonable approximation.  We note that if a sufficiently thin shell is used, the small number of cells contained within it could lead to noisy results.  However, despite using very thin shells (only 100pc thick, comparable to the width of a single cell), the resulting profiles are qualitatively quite smooth, and the results do not depend upon shell thickness.  
In the smooth-accretion simulation (dashed lines) we find that the inflow rate is nearly an order of magnitude stronger than the outflow rate (except at $< 1$ kpc scales where inflow and outflow are comparable).  The exception to this is when a galaxy merger occurs, which provides a localized spike in the inflow rate, often with a corresponding, though much weaker, spike in the ouflow rate due to a gaseous component of the infalling galaxy with velocity dispersion or circular velocity larger than the rate of infall.  Excluding the effect of incoming galaxy mergers, the inflow rate remains relatively constant outside $\sim 4$ kpc scales, below which there is often an increase in both the inflow and outflow rates. In contrast, the clumpy-accretion model can have outflow rates comparable to or higher than the inflow rates if the black hole is large enough (by $z \sim 8$ for this black hole), and the outflows extend out to large radii.

The lack of decrease in outflow rate beyond $\sim 4$ kpc suggests two things.  First, that the majority of the outflowing gas that reaches $\sim 4$ kpc tends to be at or above the escape velocity of the host galaxy (shown to be correct in Figure \ref{fig:radial_properties_67}), and second it suggests that the majority of outflowing gas that reaches $\sim 4$ kpc is able to continue outward without significant retardation by its environment.  This is consistent with Figure \ref{fig:projection_67}, which shows that the hot gas tends to expand out of the plane, thereby avoiding the dense in-plane gas that can impede the gas flow.  We investigate this directional dependence of the outflows in Section \ref{sec:in_vs_out_of_plane}.
We also note that the incoming galaxy (seen as a spike in the inflow rate in each simulation) is notably delayed in the clumpy accretion run.  This delay is likely due to the hotter gas environment through which it passes.  Since the circumgalactic gas tends to have higher outward velocities, the increased ram pressure is able to more efficiently slow the incoming galactic gas.

\begin{figure*}
\centering
\subfigure{
\includegraphics[width=8cm]{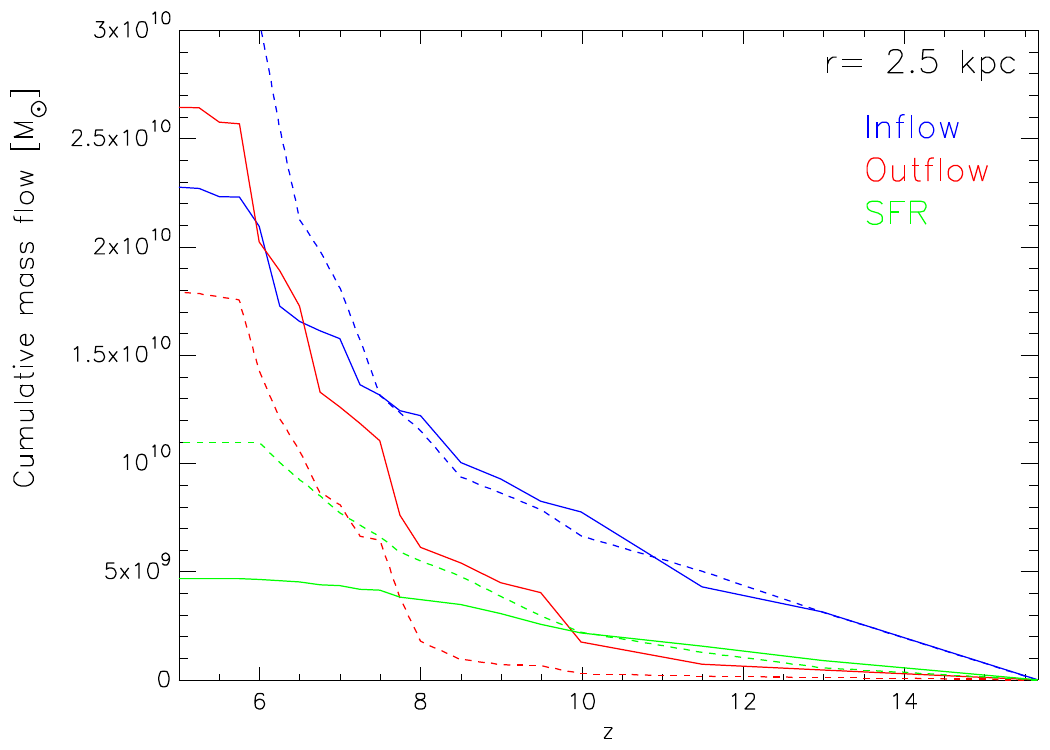}
}
\subfigure{
\includegraphics[width=8cm]{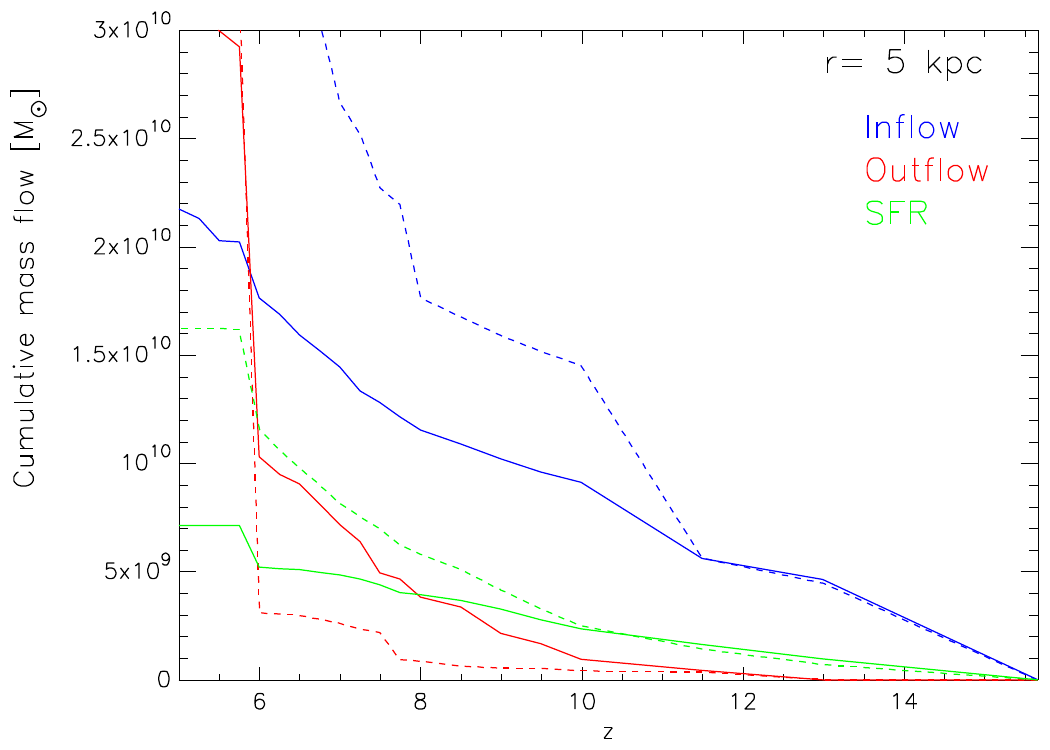}
}
\subfigure{
\includegraphics[width=8cm]{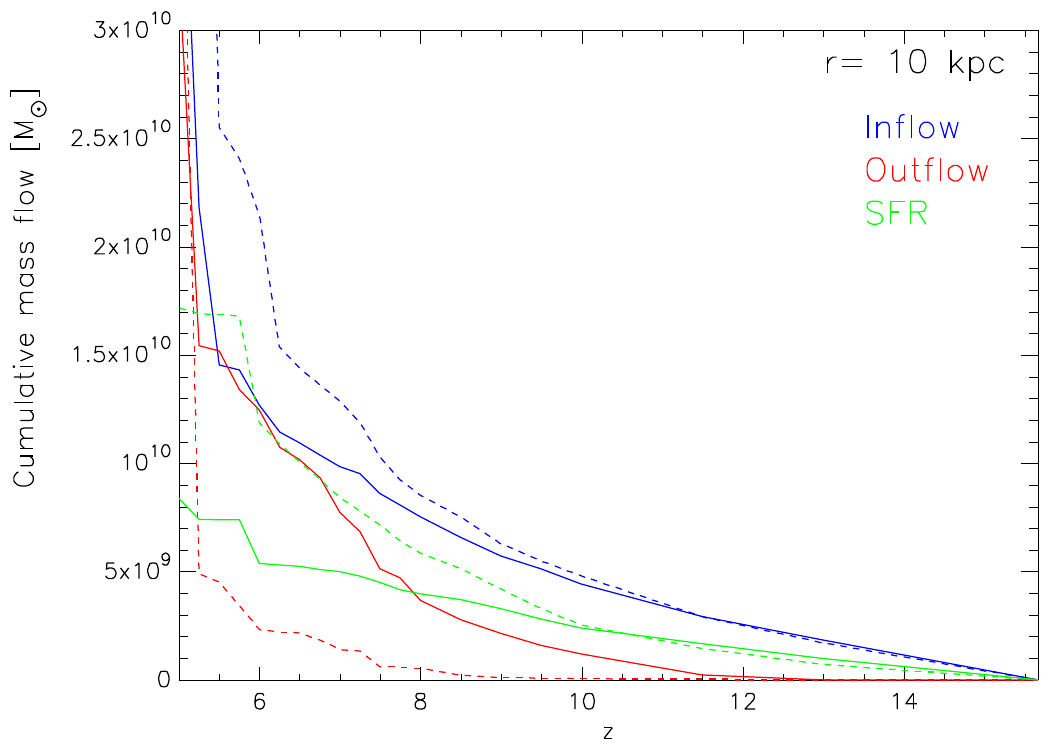}
}
\subfigure{
\includegraphics[width=8cm]{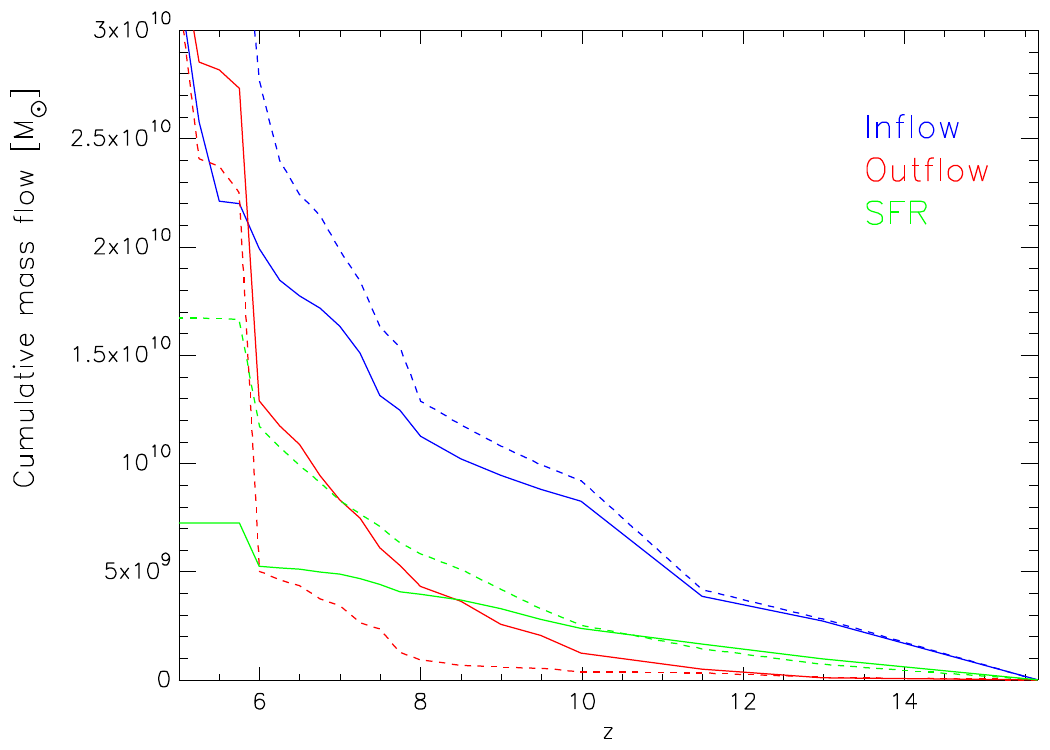}
}
\caption{Cumulative gas inflow (blue) and outflow (red) through spherical shells at 2.5kpc, 5kpc, and 10kpc (each 1 kpc thick), and a shell spanning 2-10 kpc as functions of redshift.  Also shown is the cumulative SFR (green) in the spherical region interior to the shells.  Solid lines show the rates for the clumpy-accretion simulation; dashed lines show the rates for the smooth accretion simulation.  Clumpy accretion expels more gas and suppresses both gas inflow and star formation.}
\label{fig:cumulative_gas_flow}
\end{figure*}

Although having much stronger outflow rates, Figure \ref{fig:radial_flow_rates_67} shows that the clumpy accretion run has a generally comparable inflow rate outside the innermost region to that of the smooth accretion run.  To investigate the long-term gas inflow onto the galaxy, in Figure \ref{fig:cumulative_gas_flow} we plot the cumulative gas inflow (blue) and outflow (red) through spherical shells surrounding the central galactic region for both accretion models.  This cumulative flow rate is calculated using the instantaneous flow rate at each snapshot, and assuming this rate remains constant until the next snapshot is reached.  To avoid having a single thin shell with an unusually high flow rate due to an infalling clump, we take the average flow rate through 10 shells, each 100pc thick.  
We show these cumulative curves at radii of 2.5 kpc (top left), 5 kpc (top right), and 10 kpc (bottom left), and a thick-shell curve for flow averaged across all shells between 2 and 10 kpc (bottom right).  

Considering the outflowing gas in the clumpy accretion model (solid red lines), we see that there is significantly more outflow at 2.5 kpc than at 5 kpc, since some of that gas is slowed down by the gas in the galactic disk.  At 5 and 10 kpc, however, we find similar outflow rates across cosmic time, confirming that the bulk of the outflowing gas beyond $\sim 3$ kpc continues to at least 10 kpc without significant deceleration, consistent with the instantaneous outflow rates in Figure \ref{fig:radial_flow_rates_67}. In contrast to this, the smooth-accretion model (dashed red line) shows a continued decrease in outflowing gas mass out to larger radii.  This is expected, since the much lower outflow velocities (see Figure \ref{fig:radial_flow_rates_67}) mean that much less gas from the central region where AGN-driven outflows originate is capable of escaping the potential well, and thus we see the decrease in expelled gas at higher radii.  

We also show the cumulative gas infall onto the galaxy (blue), where we again find significant differences between the clumpy- and smooth- accretion models.  At early times (prior to z$\sim$8), we find the gas mass accreted onto the host galaxy is consistent between the two models.  This is expected since at early times, the AGN feedback should be insufficient to affect the inflowing gas.  Once the black hole is massive enough, however, we note that not only does the clumpy-accretion model provide much stronger outflows, it also substantially suppresses the inflow of gas onto the galaxy, which we see beginning at $z \sim 8$ for this galaxy.  Note that at 5kpc it appears to start much earlier, but this is due to a high-inflow rate caused by an incoming galaxy in a single snapshot.  Ignoring the jump caused by this incoming merger, we again see the increased inflow in the smooth accretion case start at $z \sim 8$, also at 5kpc.  
This suppression of inflowing gas correlates directly with the onset of self-regulated growth (Figure \ref{fig:bh_growth}), suggesting that the regulation of black hole growth is correlated not only with expelling gas from the central region, but also limiting the replenishment of this reservoir through inflowing gas.

\begin{figure*}
\centering
\includegraphics[width=17cm]{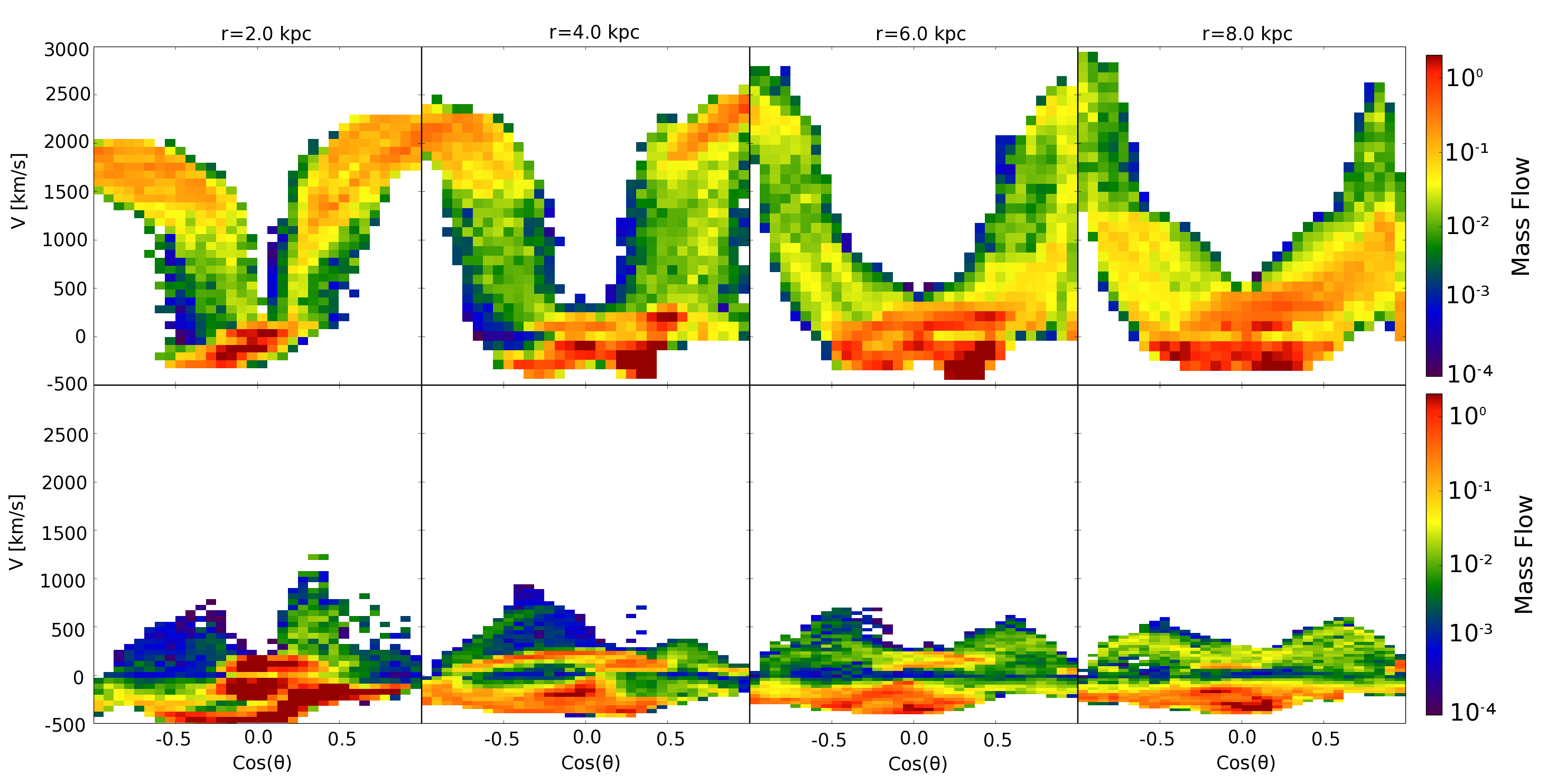}
\caption{Mass flow rate as a function of radial velocity and polar angle for spherical shells at radii 2, 4, 6, 8 kpc (columns) for the clumpy-accretion run (top) and the smooth-accretion run (bottom) at z=7.65.  Outflows in the clumpy accretion model are directed perpendicular to the galactic plane, particularly at larger radii.  }
\label{fig:velocity_angle_67}
\end{figure*}

In addition to the flow rates through spherical shells, Figure \ref{fig:cumulative_gas_flow} shows the cumulative SFR (green lines).  Because of the localization of SFR to the high density regions, we consider star formation within the spherical region interior to the given radius, rather than within a thin shell at the radius.  From these curves we can see that although AGN-driven outflows consist of hot, diffuse gas that does not form stars, the clumpy-accretion AGN nonetheless significantly quenches star formation by nearly a factor of 2.  This appears to be in contrast to the results of \citet{GaborBournaud2014} based on isolated-galaxy simulations, who found that despite driving strong outflows, the star formation rate was minimally affected.  
However, this apparent discrepancy is due to a difference in the black hole growth phase being investigated, and accounting for this brings both results into agreement with one another.
We find that the quenching of star formation occurs only after the black hole has undergone an extended phase of Eddington limited growth, while the \citet{GaborBournaud2014} investigation used a black hole which is substantially sub-Eddington (except for the bursts due to clump accretion onto the black hole).  Compared to their $\sim 100$ Myr simulation in which the black hole only grows by $\sim 15\%$ (with an averaged Eddington fraction of only a few percent), we begin seeing suppression of star formation only after the black hole grows by an order of magnitude at Eddington, and the effect becomes strong only after growing by a factor of $\sim40$. Prior to such extended growth, we are fully consistent with \citet{GaborBournaud2014}: our AGN drives strong outflows of hot, diffuse gas, entraining minimal high-density gas, and being directed almost entirely out of the galactic plane with no significant effect on star formation or host morphology (see Section \ref{sec:earlytime} for more details).

\subsubsection{Geometry of inflows and outflows}
\label{sec:in_vs_out_of_plane}

In Figure \ref{fig:projection_67} we saw that the hot gas driven by the black hole seemed to be strongly directed out of the plane of the galaxy, and in Figure \ref{fig:radial_flow_rates_67} we saw that the outflowing material did not significantly slow beyond $\sim 3$ kpc, again suggesting expansion away from the dense galactic gas that could impede its progress.  To investigate this directly, we compute the radial mass flow as a function of cos$(\theta)$, where $\theta$ is the angle relative to the polar axis of the galaxy.  We define the polar axis to be the mass-weighted angular momentum vector of the gas in the central 1 kpc of the galaxy, but we find that these results are not sensitive to the size of the region used to calculate this vector.  In Figure \ref{fig:velocity_angle_67} we show the distribution of gas in terms of radial velocity and cos$(\theta)$, in shells of radius of R=2, 4, 6, and 8 kpc and thicknesses of 0.2R, for both clumpy-accretion (top) and smooth-accretion (bottom),
Each pixel in $V_R$-cos$(\theta)$ is color coded by the total mass flux through the shell at the given velocity and angle.  In the smooth-accretion model, we see that the strongest outflow velocities tend to be out of the plane, but not substantially so, peaking at $\sim 30$ degrees above/below the plane, while the strongest flow rates (rather than flow velocities) tend to be at low velocity and primarily inward.  In the clumpy accretion model, however, we have a clear angular dependence on the outflowing velocity, with the strongest outflow rates being at the highest velocities, and strongly out of the plane.  Furthermore, this clear correlation between outflow velocity and polar angle grows with shell radius, confirming that the more out-of-plane the gas flows, the less it gets impeded as it travels outward.  

In contrast to the out-of-plane flows which are relatively unimpeded, the gas moving into the galactic plane is rapidly slowed, with rapid inflow spread over a larger range of $\theta$ at large radii (8 and 6 kpc) than small radii (4 and 2 kpc).  We also note that in the clumpy-accretion case, at 2 kpc there is outflowing gas directed into the plane (though not as strong as the out of plane), but this outflowing in-plane gas does not survive to 4 kpc.  This is due to the the void around the black hole (see Figure \ref{fig:projection_67}) which extends to $\sim 1-2$ kpc.  Within the void, in-plane gas flows freely, but is rapidly stopped upon reaching the high-density region.  Beyond the void, the only rapidly outflowing gas is that which was directed out of the galactic plane.

\subsection{Inflow suppression}

\begin{figure*}
\centering
\includegraphics[width=16cm]{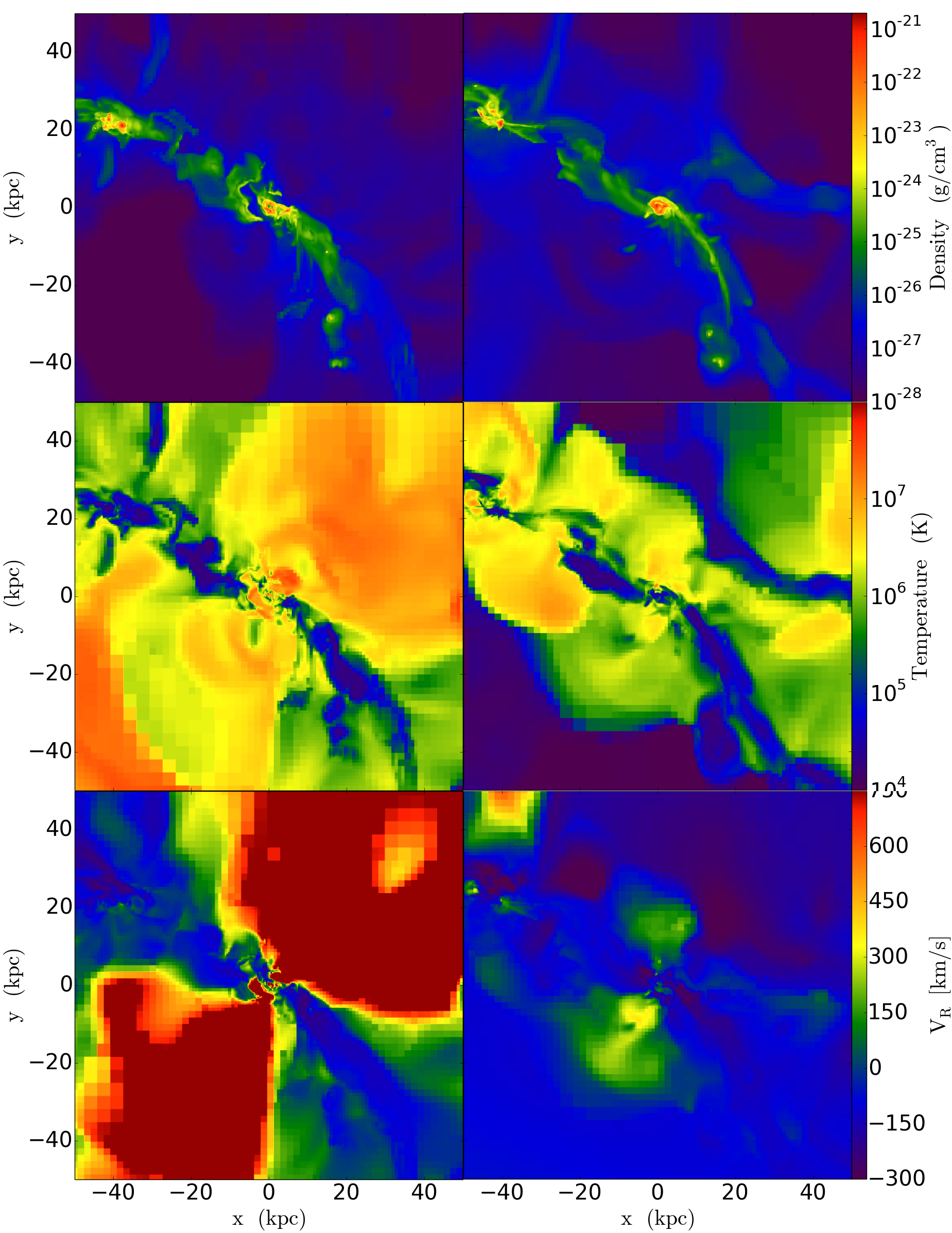}
\caption{Projection plots of our clumpy-accretion model (left) and smooth-accretion (right) models, showing gas density (top), temperature (middle), and radial velocity (bottom) at z=7.65.  Note: to more clearly show the inflow velocities, the radial velocity colorbar is limited to values within $[-300,700]$ km\/s.}
\label{fig:largescale_projection}
\end{figure*}

In Figure \ref{fig:cumulative_gas_flow} we showed that the inflowing gas is suppressed in the clumpy accretion model, showing that the AGN is able to not only drive out hot galactic gas, but affect the inflowing gas streams.  In Figure \ref{fig:largescale_projection} we show larger-scale projections of the gas density (top), temperature (middle), and radial velocity (bottom) to show the means by which the inflow is affected.  In the density projections, the smooth accretion model shows more well-defined streams which survive to small scales.  In contrast, the inflowing gas streams in the clumpy accretion model are disrupted by collisions with outflowing gas, most clearly seen by the shock front to the upper-left of the black hole.  In addition to the shocks from collisions between the inflowing and outflowing gas, the outer regions of the inflowing gas streams are stripped and blown away, and only the high-density rapidly infalling gas survives.  This is seen in the velocity map in Figure \ref{fig:largescale_projection} (bottom panels).  The colorscale only shows gas with speed below 700 km/s to more cleary show the variations among the inflowing gas.  Here we see that in the smooth accretion model, the majority of gas is flowing in toward the galaxy (blue), with gradual transition from inflowing to outflowing velocities.  In contrast, the clumpy accretion (left) shows relatively small regions where inflowing streams survive.  

\begin{figure}
\includegraphics[width=8cm]{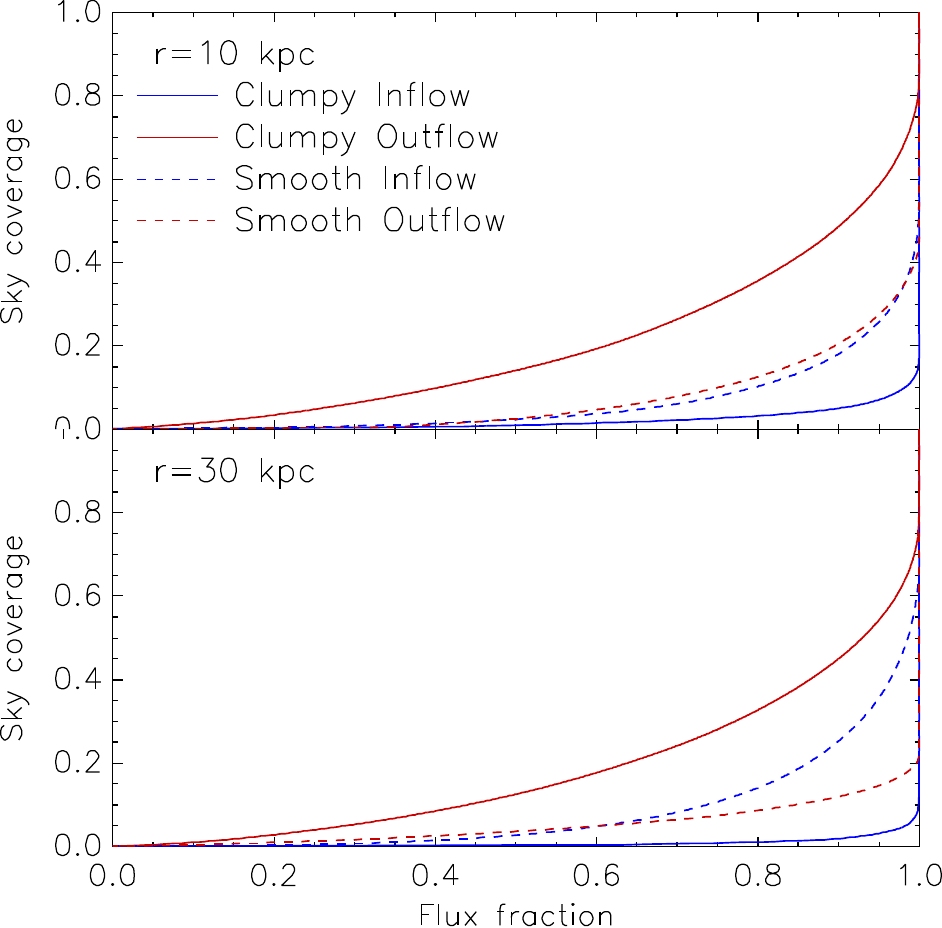}
\caption{Fraction of sky needed to include a given fraction of the total inflow (blue) and outflow (red) of the gas through a shell at 10 kpc (top) and 30 kpc (bottom) during a period of rapid growth, at z $\sim$ 8.9.  Compared to the smooth run, outflows from the clumpy accretion run are more widely distributed on the sky, while the inflows are restricted to a smaller covering fraction.  }
\label{fig:skycoverage}
\end{figure}

Furthermore, the inflowing streams completely lack the envelope of more slowly infalling gas seen in the smooth accretion model.  Instead this envelope has been stripped away, leaving a sharp transition between dense, rapidly infalling gas penetrating the rapidly outflowing gas.  This stripping effect can also be seen in Figure \ref{fig:skycoverage}, which shows the fraction of the sky needed to include a given fraction of the inflowing (blue) and outflowing (red) gas during a period of rapid black hole growth.  At $\sim 10$ kpc, the fiducial result from the smooth accretion case shows that inflowing and outflowing gas take up comparable fractions of the sky.  In the clumpy accretion model, the outflowing gas is much more widely distributed, with a corresponding compression of the inflowing gas due to the stripping effect described above.  At $\sim 30$ kpc, outflow in the fiducial run is compressed to a much smaller fraction of the sky, though note the weaker outflow here means there is very little outflowing gas.  Similarly, the clumpy accretion model again shows substantially expanded outflow comparable to the sky coverage at smaller radii, and compressed inflow.

We note that \citet{Dubois2013a} have also investigated high-redshift black hole growth and the impact on the host galaxy.  Similar to our clumpy accretion model, they found that the black hole is able to evacuate gas from the central galactic region, thereby suppressing star formation, and also reduces gas accretion onto the galaxy.  Futhermore, they also found that AGN activity can be driven by dense gas clumps migrating to the galaxy center (in addition to direct feeding by cold flows), consistent with our general model.  However, they tested low- (125pc) and high- (15pc) resolution cases, and found that the SFR history was generally consistent between the two runs (except at very high redshifts), contrary to our results presented here.  However, we note that their black hole is very efficiently fueled, starting at Eddington upon seeding, and is maintained for an extended period (growing the black hole by 2 orders of magnitude) due to efficient low angular momentum cold streams. Because these streams are sufficient to maintain Eddington starting from insertion of the black hole into the simulation, we would not expect the resolution of gas clumps to have a significant difference; rather it is in galaxies where the black hole starts at sub-Eddington accretion rates that we expect clumps to have a strong effect as shown here. 

\section{Early-time effects}
\label{sec:earlytime}

Although \citet{GaborBournaud2014} found similar outflows (see Section \ref{sec:inflow_outflow}), neither star formation nor host morphology were significantly affected, seemingly in conflict with the results presented here despite our model being calibrated using that simulation.  However, we note that those findings were based upon a short-timescale ($\sim 100$ Myr) run in which the black hole only grew $\sim 15\%$ (as shown in Figure \ref{fig:bh_growth}), and without ever having undergone an extended period of Eddington growth (the only Eddington accretion is found during the 5-10 Myr accretion events).  In contrast to this, our simulation predicts that the black hole can impact the host galaxy morphology and star formation rate after having undergone an extended Eddington phase, increasing the mass by more than an order of magnitude.

\begin{figure*}
\centering
\subfigure{
\includegraphics[width=17cm]{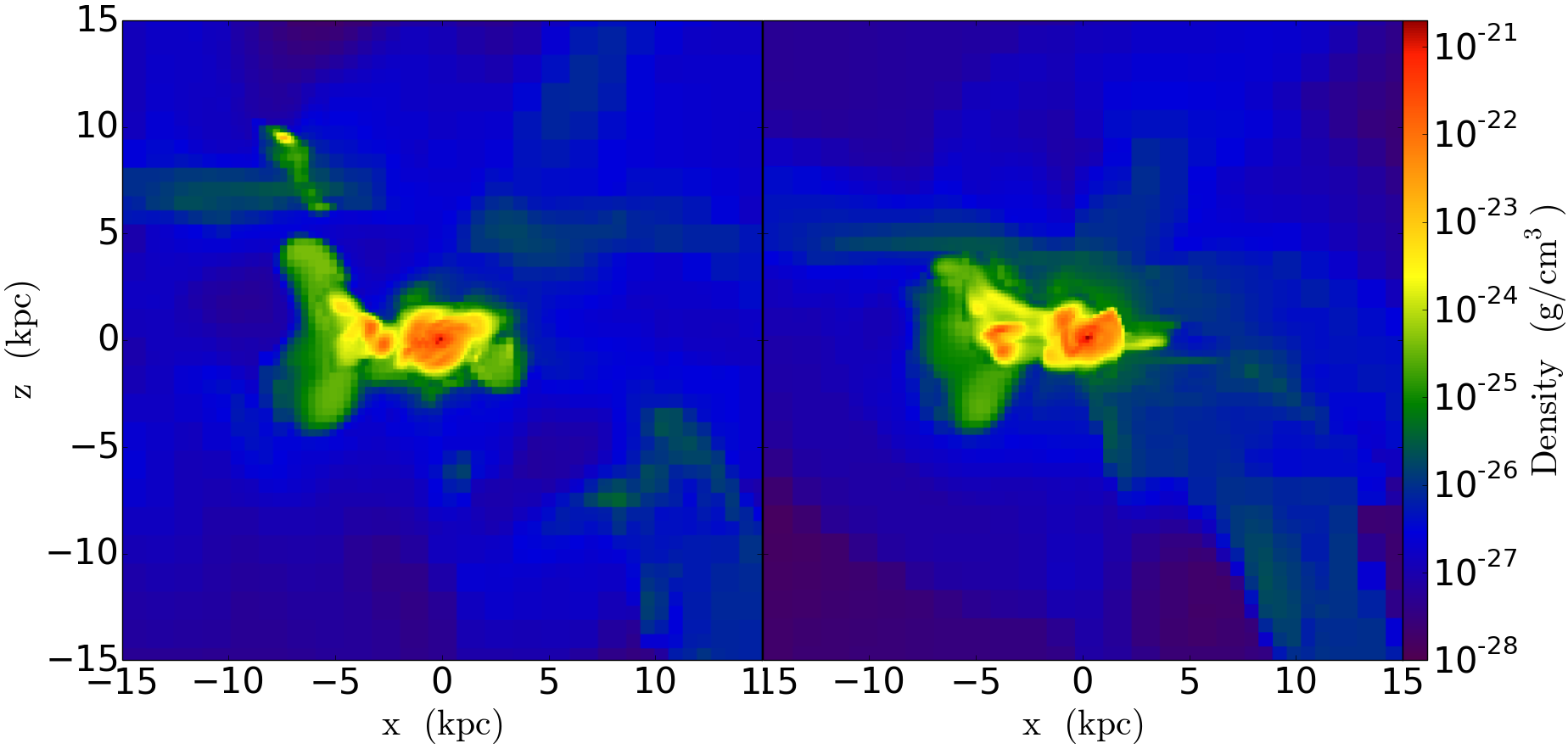}
}
\subfigure{
\includegraphics[width=17cm]{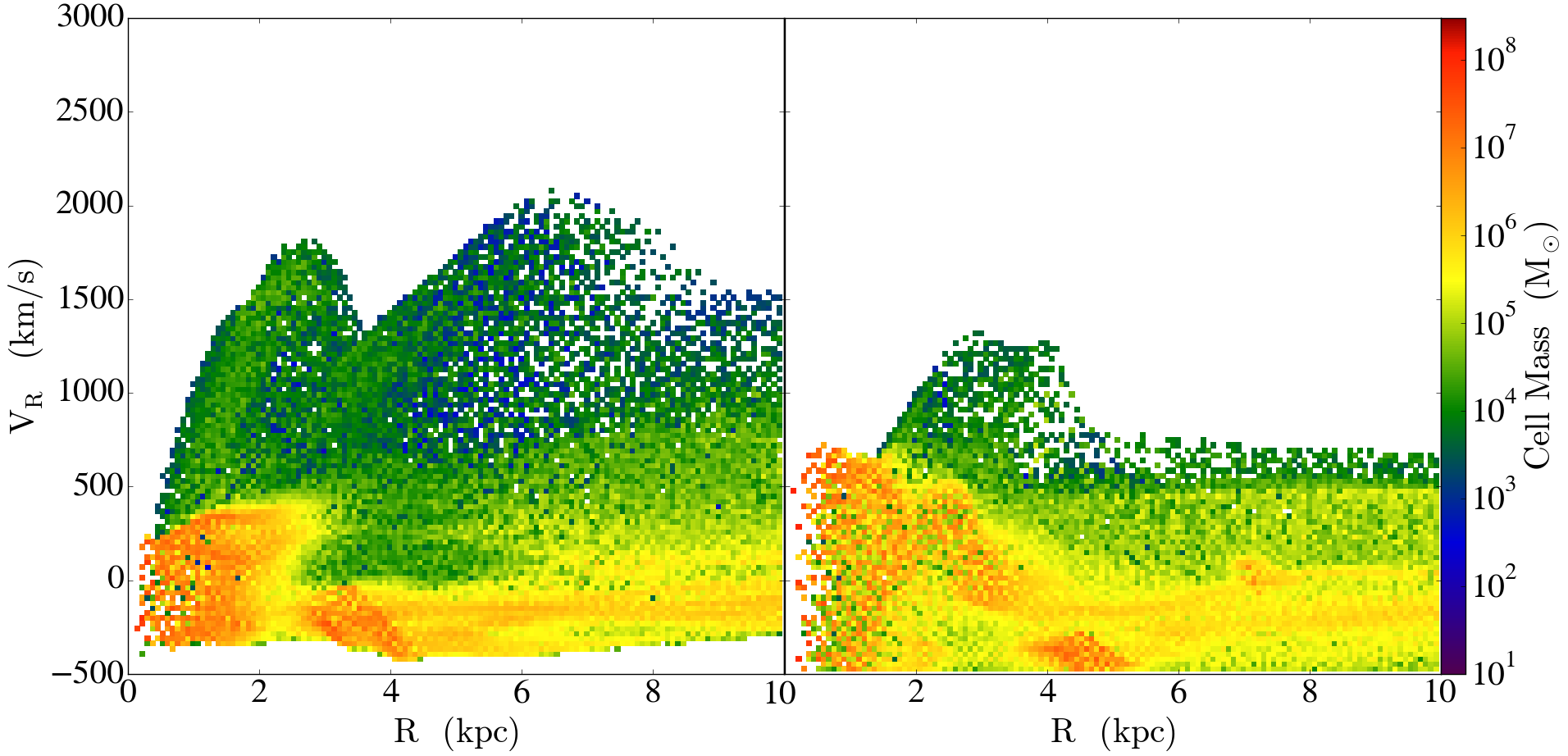}
}
\caption{Host galaxy before the black hole reaches the self-regulated regime at z $\sim$ 10.  \textit{Top:} Density map of gas in 6-kpc thick slice about the black hole.  \textit{Bottom:} Radial velocity distribution as a function of distance from galactic center.  Left hand panels show the clumpy accretion model, while the right hand panels show the smooth accretion model.  The effect of the clumpy accretion is much weaker than at later times.}
\label{fig:directcompare}
\end{figure*}

\begin{figure}
\centering
\includegraphics[width=9cm]{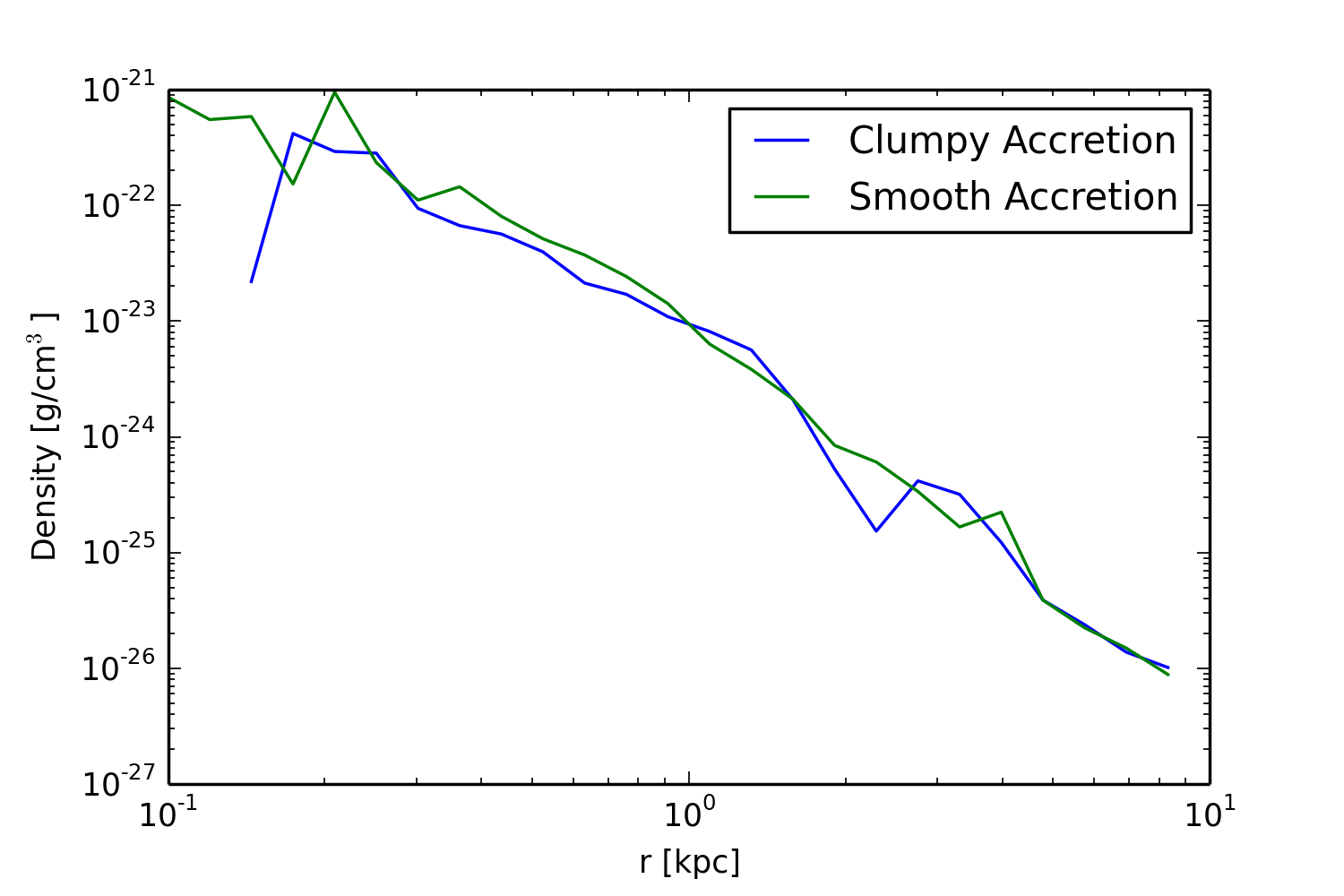}
\caption{Gas density profile for clumpy accretion model (blue) and smooth accretion model (green), prior to reaching the self-regulated regime at z $\sim$ 10.  Clumpy accretion at early time does not affect the gas density of the galaxy.}
\label{fig:directcompare_profile}
\end{figure}

To provide a more comparable case between the isolated galaxy run and our cosmological runs, we look at the host properties at an earlier time, when the black hole is smaller and has not yet approached the self-regulated regime.  Self-regulation occurs at the end of the Eddington regime, where the feedback from the black hole is strong enough to suppress its own accretion.  The onset of regulation is where we expect to find the strongest effects, which we showed in earlier sections.  To compare with the isolated galaxy, we consider the black hole and its host at $z \sim 10$, when the black hole has reached $10^6 M_\odot$ but is not yet at the self-regulated regime.  In the top panels of Figure \ref{fig:directcompare} we show the density maps of the host galaxy, finding no significant morphological effects, contrary to Figure \ref{fig:projection_67} where significant morphological differences were found for the self-regulated regime.  In the bottom panel of Figure \ref{fig:directcompare} we show the distribution of gas velocity as a function of radius, finding that the clumpy-accretion model (left) does drive significantly more gas at much higher velocities than the smooth-accretion model (right).  Thus we find that, consistent with \citet{GaborBournaud2014}, if the black hole has not yet undergone significant Eddington growth it is capable of driving strong outflows of hot, diffuse gas without having a significant effect on the rest of the host galaxy. This is further confirmed in Figure \ref{fig:cumulative_gas_flow} which shows minimal difference in high-z gas inflow or SFR between the clumpy- and smooth-accretion runs.  To quantitatively compare the morphologies, Figure \ref{fig:directcompare_profile} shows the density profile for both the clumpy- and smooth- accretion models at this early time.  The density profiles are in complete agreement, lacking the clear central void in Figure \ref{fig:density_profile_comparison} at the later, Eddington phase.  The lack of any such void shows that at early times, comparable to the conditions of \citet{GaborBournaud2014}, the black hole has not evacuated the central region, which only occurs after longer-term growth and feedback have occured.  

Thus we find that including periodic accretion of high-density gas clouds can have a strong effect on the host galaxy, but only after the black hole has grown significantly, more than an order of magnitude at $\sim$Eddington rates.  Prior to this growth, the AGN can drive rapid outflows of hot, diffuse gas without suppressing star formation or impacting the overall gas distribution of the host. 
A further investigation into the impact of periodic accretion bursts should also be performed using a high-resolution isolated galaxy, but one in which a black hole has already undergone extended Eddington growth and is approaching the self-regulated regime.  Since isolated galaxy simulations cannot be run for such extended times without running into physical limiations (e.g. exhaustion of gas supply in the absence of cosmological inflows), an alternative is to set up initial conditions in which the black hole starts in a very massive state compared to the host, but still in equilibrium.  Such simulations are beyond the scope of this paper, so we leave this investigation for a future project.

\section{Conclusions}
\label{sec:Conclusions}

We find that the increased periods of accretion caused by high-density, small scale gas clumps is an important factor in the cosmological growth of black holes, affecting both the black hole growth and the impact upon the host evolution.

\begin{itemize}

\item Inclusion of clumpy-accretion allows for a significant boost to black hole growth starting at early times.  Prior to the onset of Eddington-limited growth, although the total mass accreted during these clump phases is comparable to the total mass accreted during smooth phases, the \textit{net effect is much larger}.  Because sub-Eddington growth depends on $M_{\rm{BH}}^2$ (see Equation \ref{eqn:bondi}), the increased mass due to growth from the clump accretion also serves to increase the accretion rate during the smooth periods, reaching high-masses at much earlier times than in the absence of clumpy accretion.

\item The increased feedback in the clumpy-accretion model has a significant impact on the host morphology: The central $\sim$1 kpc region about the black hole is mostly evacuated of gas, while at larger radii ($\sim 7-8$ kpc) the gas density is higher due to the increased feedback-driven outflows.

\item In the absence of clumpy-accretion, the inflow is generally an order of magnitude stronger than the outflow beyond the innermost few kpc.  In contrast, the clumpy-accretion model has outflows $\sim 10$x stronger, comparable to the inflow rates (excluding incoming galaxy mergers).  

\item The bulk of the feedback-driven outflows are out of the plane of the galaxy.  The feedback energy is deposited isotropically, so the polar outflows are a purely environmental effect, caused by the high-density in-plane gas obstructing in-plane outflows.  This effect holds out to large radii, with a tendancy for the larger-radius outflows to be even more highly collimated.  

\item In the clumpy accretion model, AGN feedback nearly entirely halts inflow of gas on the $\sim$kpc scale, and at larger scales can suppress gas inflow by nearly a factor of two.  This suppression of inflow has two main causes: the outflows from the galaxy center directly interact with the inflowing streams and can even stop them; and more generally, the outflows strip the lower-density, lower-velocity envelope of gas around the high-density streams.

\item As a result of the stronger outflows and suppressed inflows, the SFR in the clumpy accretion case can be suppressed by as much as a factor of $\sim$2.  However, this difference only occurs after the black hole has undergone an extended period of Eddington growth, growing by at least an order of magnitude.  Prior to this extended growth, the SFR remains unaffected.

\item Most of the outflow driven by the strong AGN feedback is strong enough to exit the galaxy, without undergoing significant recycling. 

Thus we have demonstrated the importance of incorporating the effects of high-density gas clouds in cosmological simulations, and that applying a stochastic subgrid model to include them can lead to significant changes in host evolution.  Having shown the strength this periodicity can have, a more in-depth investigation is necessary to constrain the exact parameterization of the subgrid model. We emphasize that the parameters used here are based upon a single isolated galaxy simulation, and treated as if they hold universally.  Although our simulated galaxy does indeed maintain a high enough gas fraction to support our choice of baseline model, it nonetheless remains an oversimplification. This model was sufficient to demonstrate the importance that dense gas clumps (and variability in general) can have on black hole growth and the corresponding impact on host galaxy evolution, but it does not provide a statistical sample for the relative importance in large populations of black holes. Further high-resolution simulations will be needed to explore the parameter space of potential hosts to determine how the frequency and strength of incoming gas clouds depends upon various properties, including, but not limited to, host mass, gas fraction, stellar mass, disc height, merger history, etc.  With a better-constrained set of host-dependent parameters for the bursts of accretion, a full statistical analysis must be done to determine the effect on statistical samples of black holes, including possible observable signatures in the quasar luminosity function and luminosity-dependent clustering behavior.
This continuation goes beyond the scope of this paper, and will be addressed in a followup work.

\end{itemize}

\section*{Acknowledgments}

This work was supported by ISF grant 24/12, by GIF grant G-1052-104.7/2009, by a DIP grant,
by the I-CORE Program of the PBC, by ISF grant 1829/12,
by NSF grants AST-1010033 and AST-1405962, and supported from the E.C. through an ERC grant StG-257720.  Some of the simulations used in this work were performed on GENCI resources at TGCC (project 04-2192).

 \bibliographystyle{mn2e}       
 \bibliography{astrobibl}       

\end{document}